\documentclass{aa}  
\usepackage{natbib}
\usepackage{graphicx}
\usepackage{float}
\usepackage{txfonts}
\usepackage[luatex]{hyperref}
\usepackage{color}
\usepackage{xcolor}
\begin{document}

   \title{A systematic study on the properties of \\
          aromatic and aliphatic hydrocarbon dust \\
          in active galactic nuclei with AKARI near-infrared spectroscopy}

   \author{R. Katayama\inst{1}, H. Kaneda\inst{1}, T. Kokusho\inst{1}, T. Kondo\inst{1}, S. Oyabu\inst{2},
           T. Suzuki\inst{3}, and  T. Tsuchikawa\inst{1}}

   \institute{Graduate School of Science, Nagoya University, Furo-cho, Chikusa-ku, Nagoya, Aichi 464-8602, Japan\\
              \email{r.katayama@u.phys.nagoya-u.ac.jp} 
              \and
              Institute of Liberal Arts and Sciences, Tokushima University, 1-1 Minami-Jyosanjima, Tokushima-shi, Tokushima 770-8502, Japan
              \and
              Institute of Space and Astronomical Science, Japan Aerospace Exploration Agency, 3-1-1 Yoshinodai, Chuo-ku, Sagamihara
              Kanagawa 252-5210, Japan
              }

   \date{Received MM DD, YYYY; accepted MM DD, YYYY}

  \abstract 
    {Recent near- and mid-infrared (IR) observations have revealed the existence of appreciable amounts of aromatic and aliphatic hydrocarbon dust
     in the harsh environments of active galactic nuclei (AGNs), the origins of which are still under discussion.
     In this paper, we analyze the near-IR spectra of AGNs obtained with AKARI 
     in order to systematically study the properties of the aromatic and aliphatic hydrocarbon dust affected by AGN activity.
      We performed spectral fitting and spectral energy distribution fitting for our sample of 102 AGNs 
      to obtain the fluxes of the aromatic and aliphatic spectral features,
      the total IR luminosity ($L_{\rm{IR}}$), and the fractional luminosity of AGN components ($L_{\rm{AGN}}/L_{\rm{IR}}$).
      As a result, we find that $L_{\rm{aromatic}}/L_{\rm{IR}}$ is systematically lower for the AGN sample and
      especially lower for AGNs with the aliphatic feature seen in the absorption than for star-forming galaxies (SFGs),
      while $L_{\rm{aliphatic}}/L_{\rm{aromatic}}$ is systematically higher for the AGN sample than for the SFG sample,
      increasing with AGN activity indicated by $L_{\rm{AGN}}/L_{\rm{IR}}$.
      In addition, the profiles of the aliphatic emission features of the AGN sample are significantly different from those of the SFG sample 
      in that the AGNs have systematically stronger feature intensities at longer wavelengths.
      We conclude that both aromatic and aliphatic hydrocarbon dust are likely of circumnuclear origin, 
      suggesting that a significant amount of the aliphatic hydrocarbon dust may come from a new population 
      created through processes such as the shattering of large carbonaceous grains by AGN outflows.}

   \titlerunning{A systematic study on the properties of aromatic and aliphatic hydrocarbon dust in AGN with AKARI}
   \authorrunning{R. Katayama et al.}  
   \maketitle

\section{Introduction\label{sect_intro}}
There are various types of dust in the interstellar medium (ISM), 
 some of which are observed in the infrared (IR) spectral region with characteristic emission and/or absorption features.
As typical examples, aromatic and aliphatic hydrocarbon dust can be probed 
 by several spectral features in the near- and mid-IR ranges.
Regarding the aromatic hydrocarbon dust, so-called polycyclic aromatic hydrocarbons (PAHs),
 which are known to exist almost ubiquitously in the ISM, 
 possess planar structures composed of hexagonal rings with 50--1000 carbon atoms (e.g., \citealt{Allamandola1985}; \citealt{Tielens2008}).
The conspicuous emission features of the aromatic hydrocarbons are observed at
 3.3, 6.2, 7.7, 8.6, 11.3, 12.7, and 17 $\mu$m, which are produced by 
 stretch, vibrational, or bending modes of C$-$C or C$-$H bonds (e.g., \citealt{Draine2007}).
Since these features originate from the excitation of PAHs by absorbing photons from young stars in many cases (e.g., \citealt{Tielens2005}),
 the PAH emission has often been used as an indicator of recent star formation (SF) activity.
On the other hand, aliphatic hydrocarbon features are observed at 3.4, 6.9, and 7.3 $\mu$m (e.g., \citealt{Dartois2007}),
 which have multiple spectral substructures produced by 
 the C$-$H stretching and bending modes of methyl ($-\mathrm{CH_3}$), methylene ($-\mathrm{CH_2}$), or C$-$H groups 
 (e.g., \citealt{Duley1983}; \citealt{Kwok2007}; \citealt{Dartois2007}).
For the aliphatic hydrocarbons, their spectral features are found not only
 in the emission (e.g., \citealt{Yamagishi2012}) but also in the absorption (e.g., \citealt{Imanishi2010});
 historically, the latter form has been detected more often from Galactic targets (e.g., \citealt{Sandford1995}).
Since the aliphatic absorption feature is likely to be attributed to hydrogenated amorphous carbon grains (a-C:H; e.g., \citealt{Dartois2007}),
which are different from the likely carrier of the aliphatic emission feature,
we denote the aliphatic absorption feature as a-C:H to distinguish between the aliphatic emission and absorption throughout the paper.

Active galactic nuclei (AGNs) are known to have a significant impact on the properties of 
  the circumnuclear medium in a host galaxy.
In general, the PAH emission is suppressed or absent in a galaxy hosting a powerful AGN
 (e.g., \citealt{Smith2007}; \citealt{LaMassa2012}; \citealt{Esparza-Arredondo2018}; \citealt{Sajina2022}),
 although it is unclear whether it is because the PAH emission is diluted by strong IR continuum emission (e.g., \citealt{Alonso-Herrero2014}) or
 because PAHs are destroyed in the severe environment of the circumnuclear region (e.g., \citealt{Roche1991}; \citealt{Voit1992}).
Nevertheless, several observations have significantly detected the PAH emission features even in powerful AGNs.
\cite{Alonso-Herrero2014} suggested that PAHs may survive in the circumnuclear region 
 because the dense cold gas plays a role in shielding PAHs from the hard radiation field of the AGN.
\cite{Jensen2017} found that the 11.3 $\mu$m PAH emission intensity increases toward the nuclei, 
 suggesting the possibility that photons from the AGNs can directly excite PAHs within a kiloparsec from the nuclei. 

Since the hydrocarbon features are expected to be affected by the surrounding interstellar conditions,
 several studies have discussed the relationship between the fraction of the aliphatic to the aromatic feature intensities and the galactic environment.
In the nearby starburst galaxy M82, \cite{Yamagishi2012} found that 
 the aliphatic to aromatic flux ratio in the halo region increases with the distance from the galactic center,
 suggesting the production of small carbonaceous dust, which carries the 3.4 $\mu$m aliphatic emission feature
 through the shattering of larger carbonaceous grains with the galactic superwinds of M82.
A systematic study of star forming galaxies (SFGs) in \cite{Kondo2024} has shown that
 the aliphatic to aromatic ratios anticorrelate with the interstellar radiation field strength
 and that galaxies with extremely low aliphatic-to-aromatic ratios are dominated by merger galaxies.
The authors concluded that the strong radiation field due to intense star formation and/or mechanical shocks by the galactic merger destroyed the hydrocarbon dust,
 preferentially for the aliphatic hydrocarbons whose bonds are chemically weaker than the aromatic bonds.

Recently, the James Webb Space Telescope (JWST; \citealt{Gardner2023}) has revealed detailed properties of PAHs in the multiphase ISM of nearby galaxies
 at unprecedented spatial resolution and sensitivity in spectroscopy (e.g., \citealt{Rigopoulou2024}) and PAH-band imaging (e.g., \citealt{Sandstrom2023}).
As for studies on PAHs in nearby AGNs, for example,
 \cite{Lai2023} found that the aliphatic to aromatic ratios in the galactic nuclear region
 of the nearby Seyfert galaxy NGC 7469 are 
 significantly lower than those in other regions far from the nucleus, suggesting the photodestruction of 
PAHs in the nuclear regions.
 \cite{Garcia2022b} investigated the PAH properties of three Seyfert galaxies.
The authors indicatated that most of the emission originates from neutral PAHs, 
 possibly due to the effective shielding of the PAHs by high hydrogen column densities,
 and thus insufficient contribution of the AGN activity to PAH destruction,
  or the selective destruction of ionized PAHs in AGN-dominated environments.
Through a systematic study of JWST/MIRI spectra of local luminous infrared galaxies,
 \cite{Rigopoulou2024} found that the fractions of the ionized PAHs in SFGs are larger than those in AGNs,
  while the neutral PAHs are dominant in conditions under hard radiation fields such as those in the nuclear regions of AGNs and their outflows.
Furthermore, \cite{Garca-Bernete2024} compared the properties of PAHs in the AGN-dominated regions along the outflow direction
 with those in the SF regions and the nuclear regions in local AGNs. 
They reported that the AGN activity might affect the PAH population on a kiloparsec scale;
 the fact that the fraction of the ionized PAHs in the outflow regions are lower than that in the SF regions
 suggests that the ionized PAHs are preferentially destroyed in the harsh environments of AGNs.
From the observations of type-2 quasars in the central kiloparsec,
\cite{Ramos2025} report that the PAH-derived SF rates are significantly lower than those derived from optical spectra 
due to the combination of destruction by the strong radiation of the central AGNs
and relatively low gas column densities.

In this paper, we analyze the near-IR spectra of galaxies harboring AGNs obtained with 
 the Infrared Camera (IRC; \citealt{Onaka2007}) on board AKARI \citep{Murakami2007}
 in order to systematically study the properties of the aliphatic and aromatic hydrocarbon features affected by AGN activity.

 \begin{figure*}[t]
  \centering
   \includegraphics[width=5.5cm]{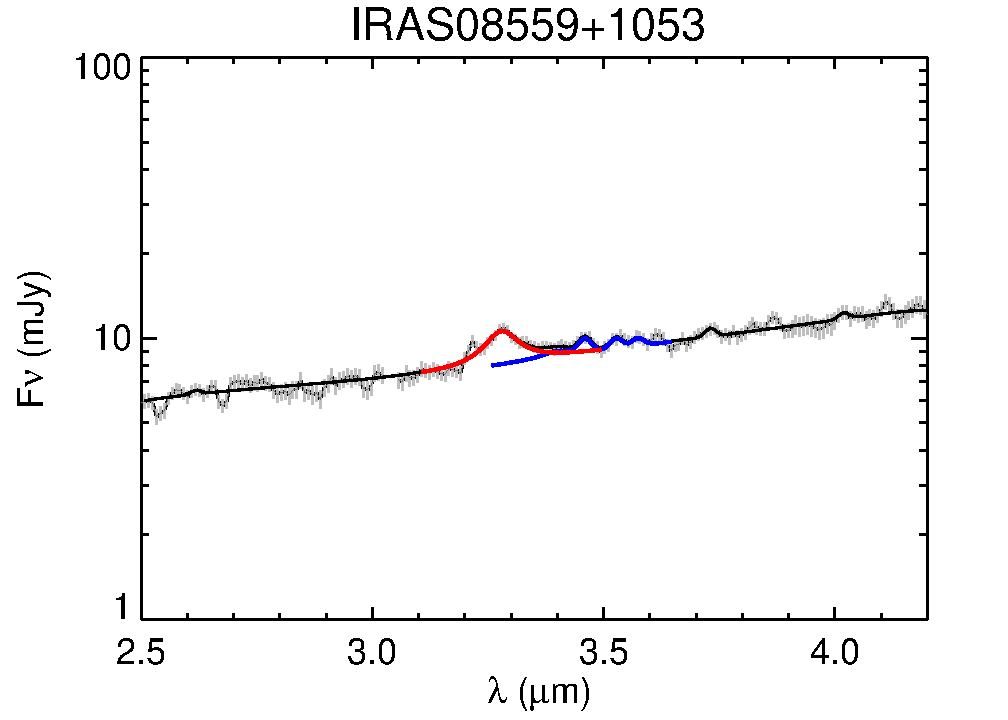} 
   \includegraphics[width=5.5cm]{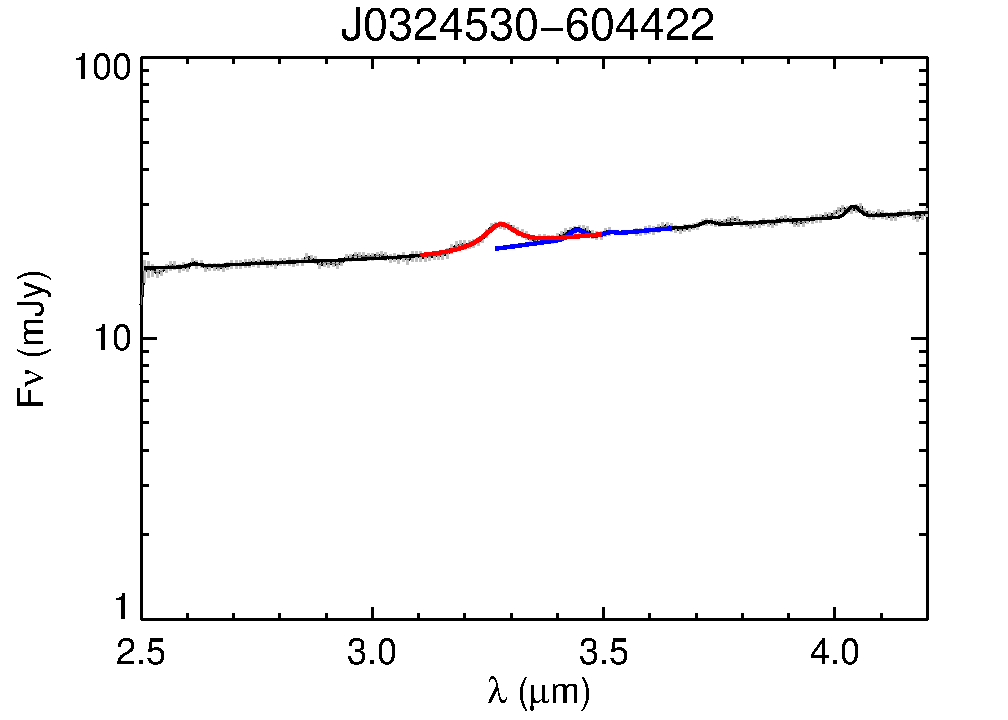}
   \includegraphics[width=5.5cm]{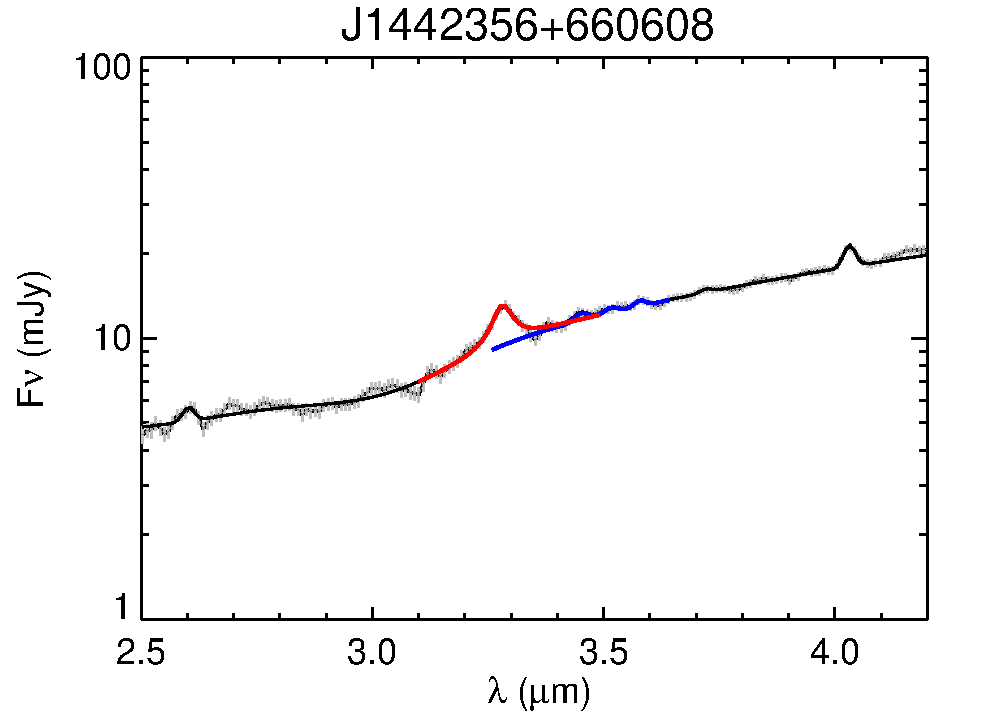}
   \includegraphics[width=5.5cm]{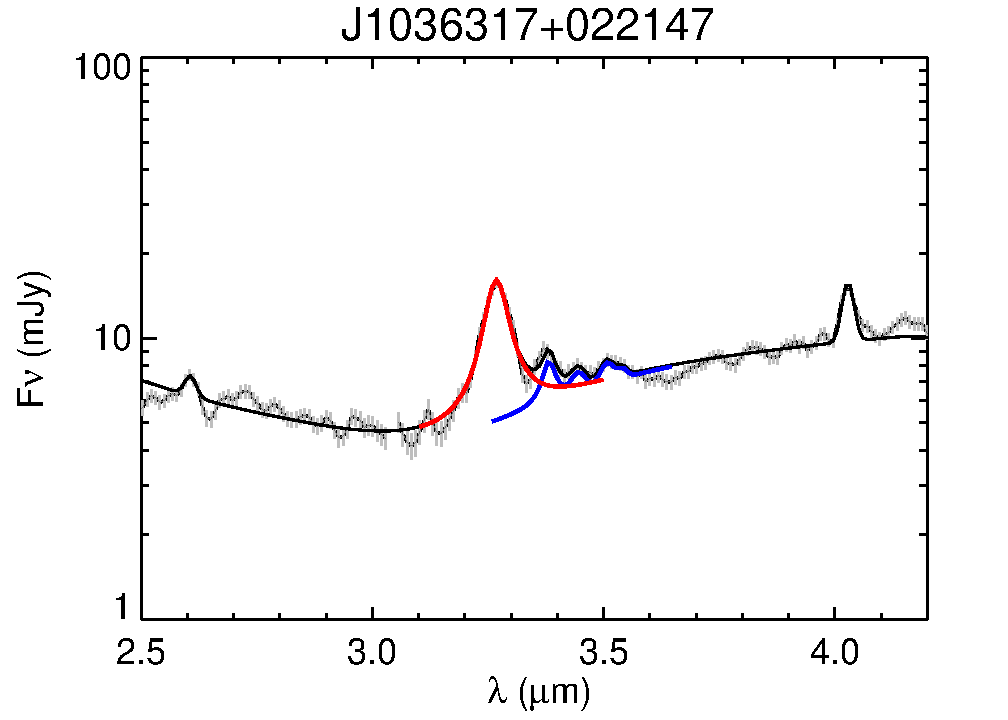} 
   \includegraphics[width=5.5cm]{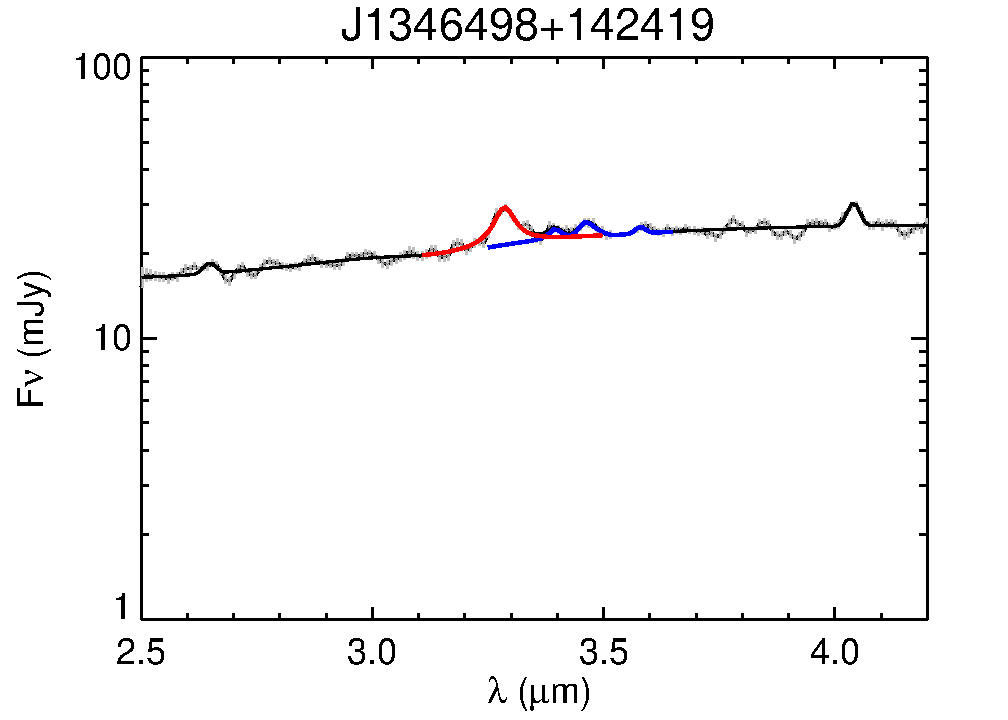}
   \includegraphics[width=5.5cm]{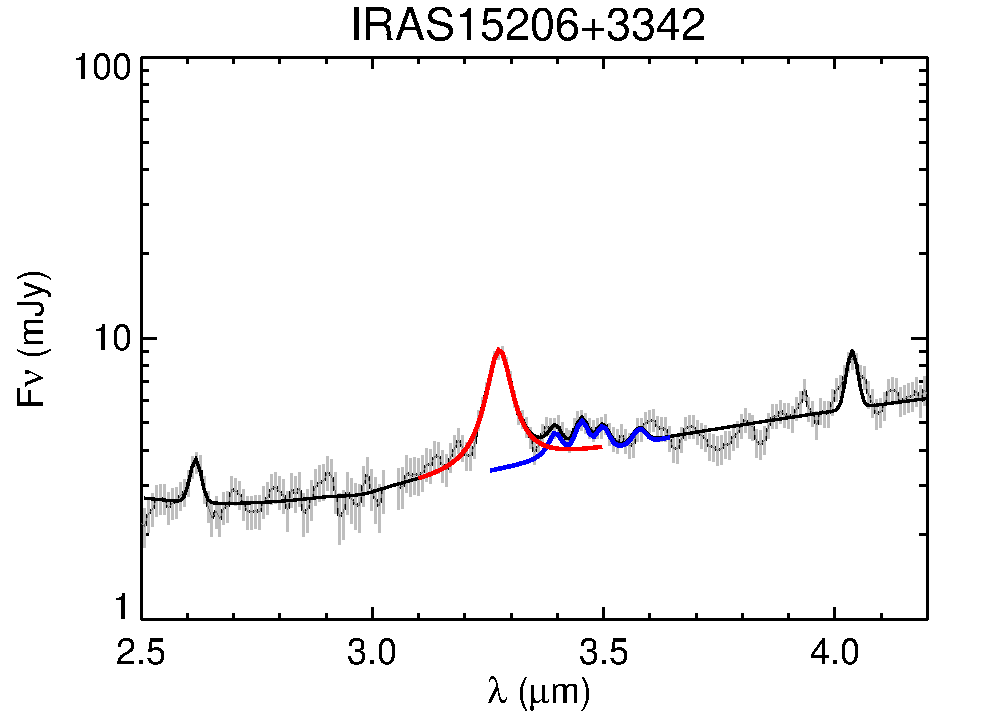}
    
   \includegraphics[width=5.5cm]{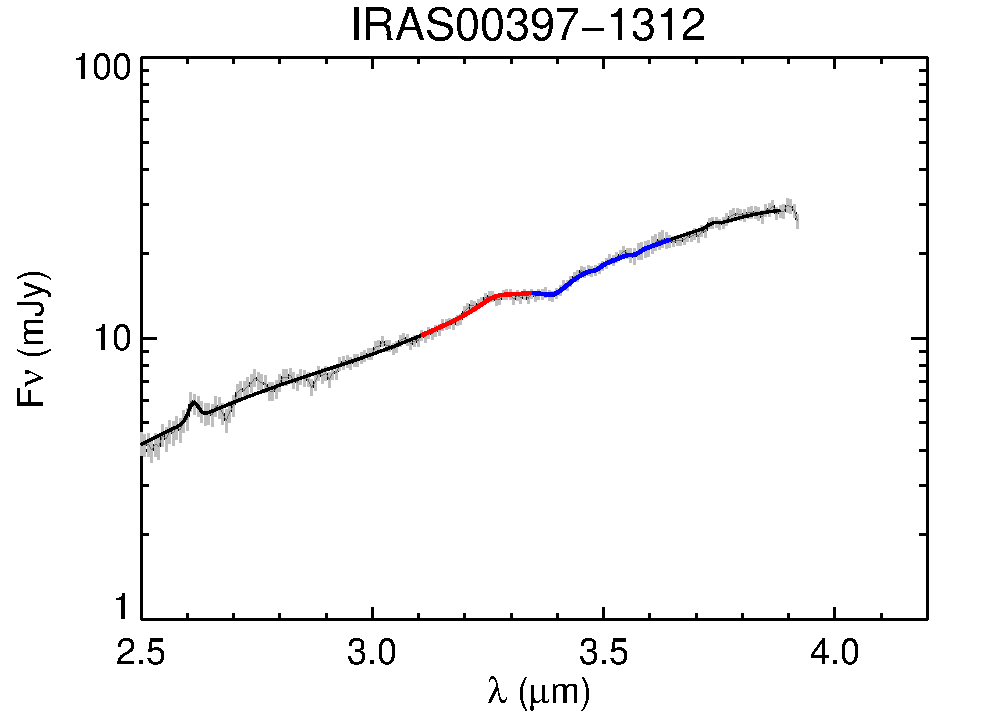} 
   \includegraphics[width=5.5cm]{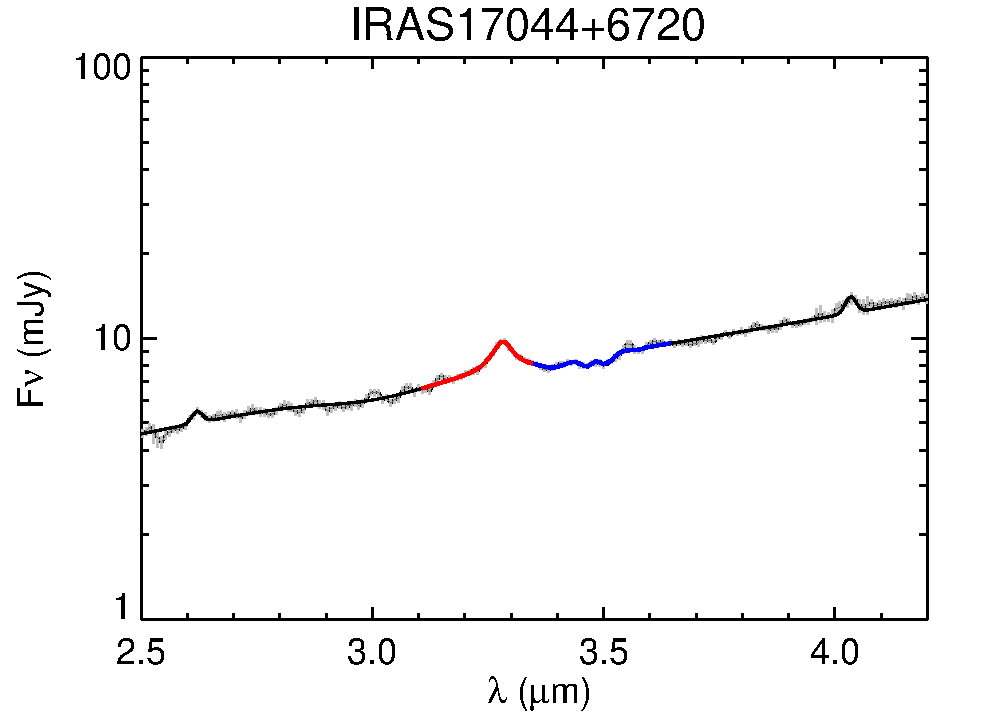} 
   \includegraphics[width=5.5cm]{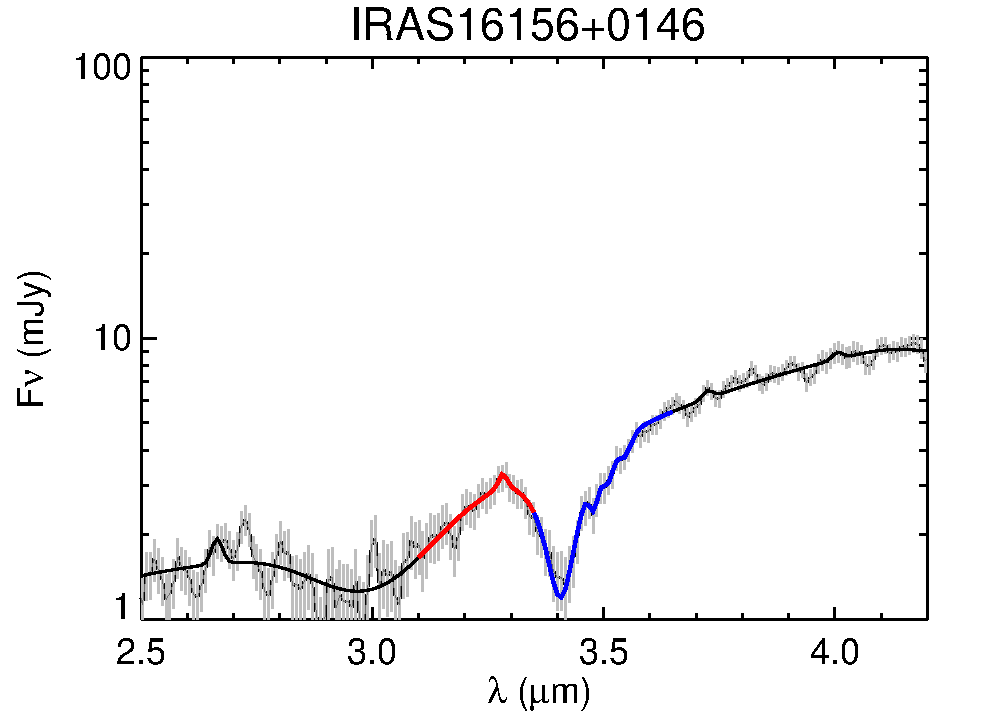}
   \includegraphics[width=5.5cm]{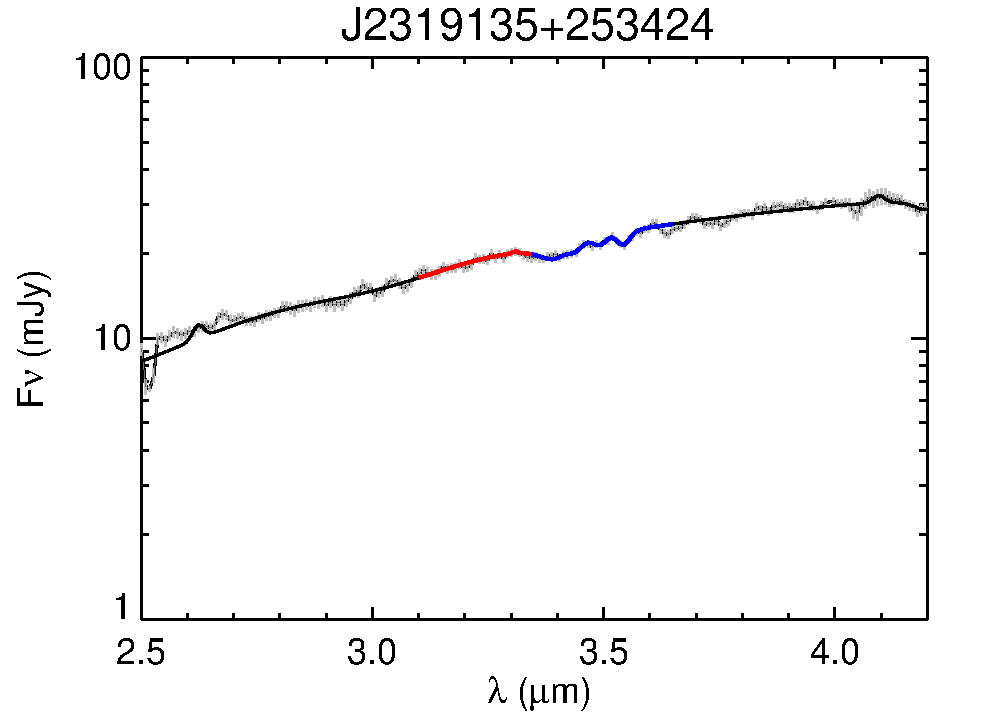}
   \includegraphics[width=5.5cm]{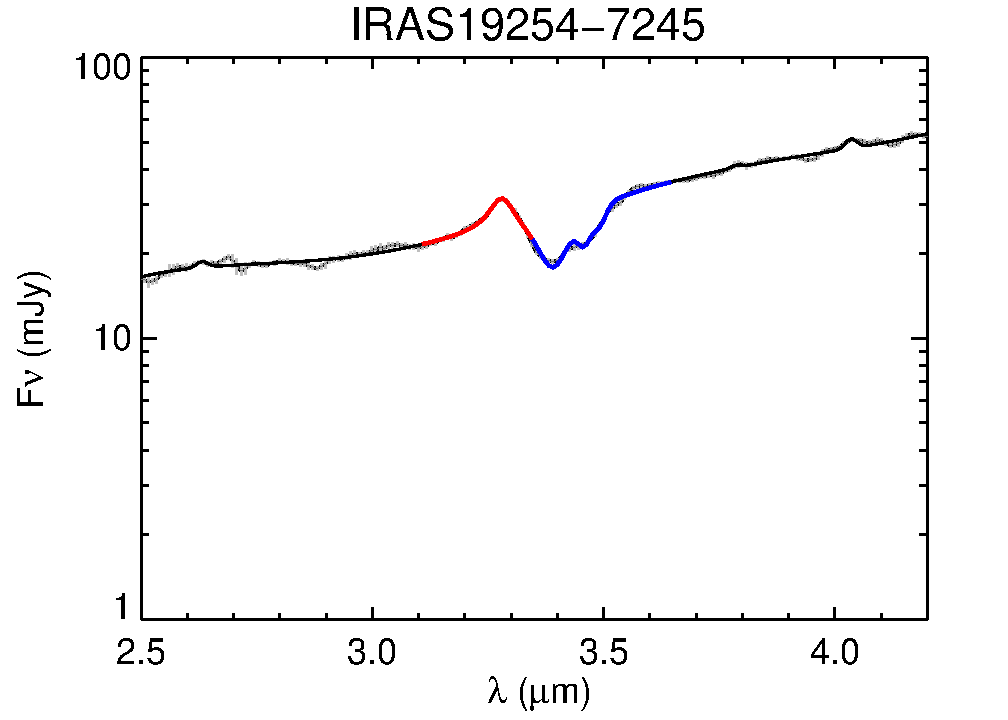} 
   \includegraphics[width=5.5cm]{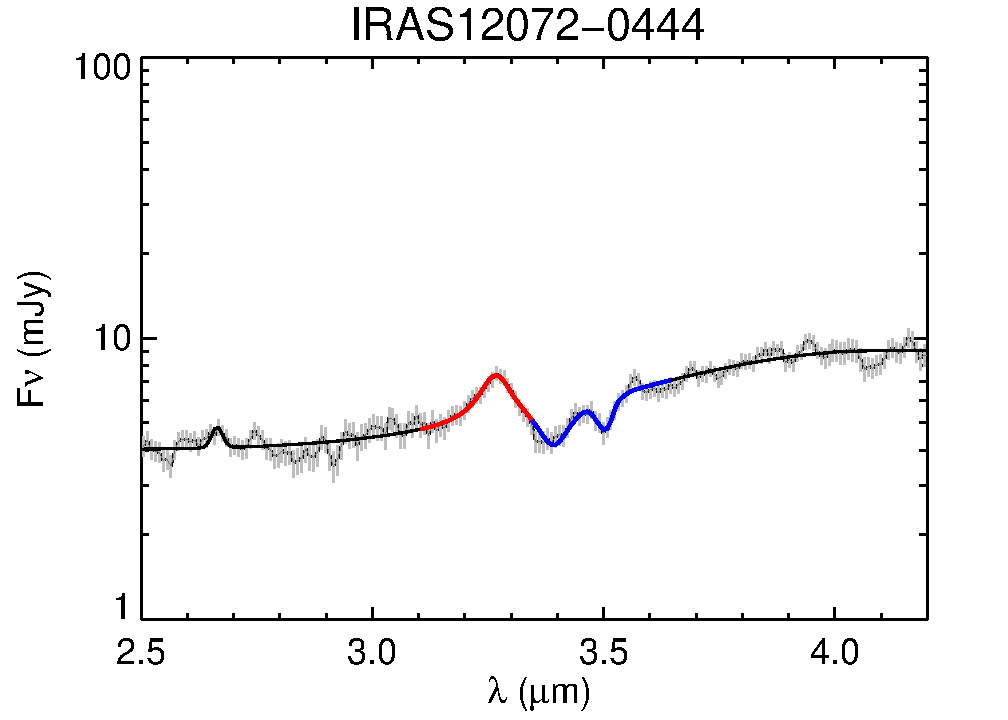}
  \caption{Examples of the spectral fitting results for the AKARI/IRC rest frame 2.5--4.2 $\mu$m spectra of
               AGN-dominated and AGN-SF-composite galaxies.
              The top two rows show examples of the cases where the aliphatic hydrocarbon features are detected in the emission,
               while the bottom two rows show those detected in the absorption.
              A black line shows the total model spectrum,
               while red and blue lines correspond to the aromatic and aliphatic hydrocarbon features, respectively.}
\label{fig:specfit_sample}
\end{figure*}

\section{Observation and data analysis}
\subsection{Sample selection}
For the present study, we selected the same samples as used in \cite{Kondo2024}, 
 which consist of the galaxies based on the two AKARI mission programs, 
 Mid-infrared Search for Active Galactic Nuclei (MSAGN; \citealt{Oyabu2011}) and
 Evolution of ULIRGs (ultra-luminous infrared galaxies, $L_{\rm{IR}} > 10^{12} \rm{L_{\rm{\odot}}}$) and AGNs (AGNUL; \citealt{Imanishi2008, Imanishi2010}). 
To exclude the normal stars and quiescent galaxies from the original samples of the above two programs,
 \cite{Kondo2024} set the sample selection criterion as $F$(9 or 18 $\mu$m)/$F(K_{\mathrm{s}})>2$,
 where $F$(9 or 18 $\mu$m) and $F(K_{\mathrm{s}})$ are the flux densities 
 at the 9 or 18 $\mu$m band and the $K_{\mathrm{s}}$ band derived from the photometry of the AKARI and 2MASS data, respectively.
Based on the criterion, they selected 230 mid-IR excess galaxies,
 which were regarded as IR galaxies owing to AGNs and/or active SF.
Each sample galaxy was classified as AGN-dominated, AGN-SF-composite, or pure-SF galaxies,
 the numbers of which were 68, 24, and 138, respectively,
 based on the criteria using the equivalent width of the 3.3 $\mu$m PAH emission ($EW_{\mathrm{PAH\, 3.3\, \mu m}}$)
 and the near-IR continuum slope as described in \cite{Kondo2024}. 
However, these criteria do not account for obscured AGNs such as those hosting compact obscured nuclei (CONs), 
 which can display an $EW_{\mathrm{PAH\, 3.3\, \mu m}}$ that is as large as the pure-SF galaxies (e.g., \citealt{Garcia2025}).
 On the other hand, the criterion using the equivalent width of the PAH emission at 6.2 $\mu$m ($EW_{\mathrm{PAH\, 6.2\, \mu m}}$) 
  enables more robust diagnosis.
 Therefore, we adopted the additional criterion of $EW_{\mathrm{PAH\, 6.2\, \mu m}} < 270$ nm \citep{Stierwalt2013}
  and reclassified the pure-SF galaxy sample using the Spitzer/IRS archival spectra.
 As a result, 17 galaxies were found to be candidates for CONs.
Notably, of the 17 galaxies, four were identified as CONs in previous studies (\citealt{Falstad2021}; \citealt{Garcia2022}; \citealt{Donnan2023}),
  and therefore we added them to the sample as AGN-SF-composite galaxies.
Furthermore, six sub-IRG AGNs (e.g., Seyfert galaxies) taken from the AKARI archives were added to the sample
 to improve the statistics at lower luminosities.
In this paper, we focus on 74 AGN-dominated and 28 AGN-SF-composite galaxies (i.e., the total 102 AGNs).
  \begin{figure*}[h]
    \centering
    \includegraphics[width=5.5cm]{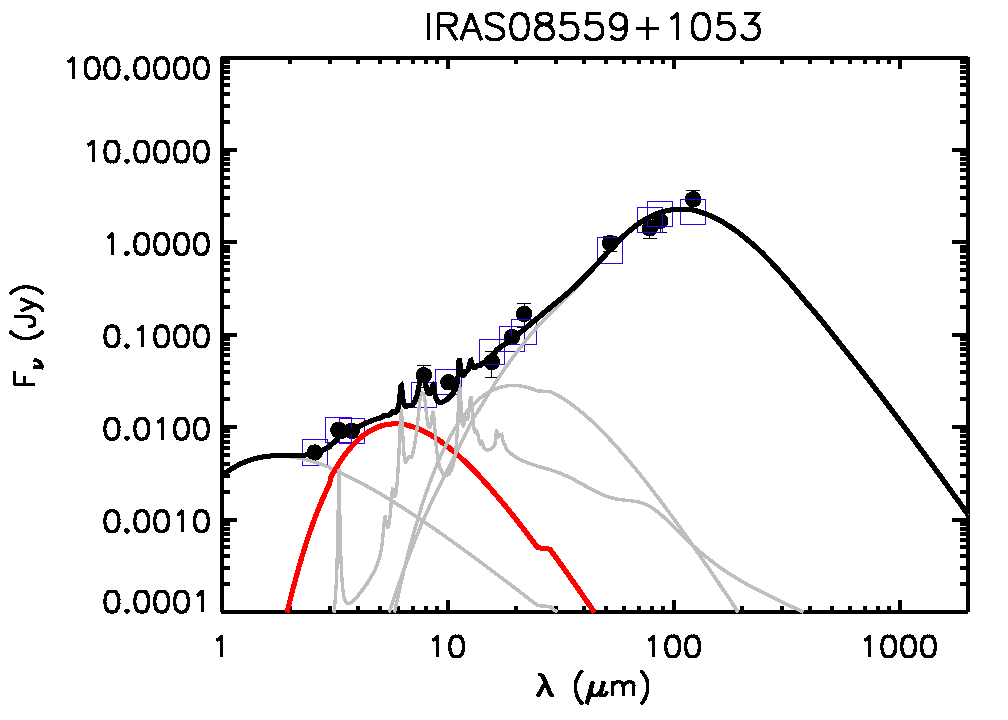}  
    \includegraphics[width=5.5cm]{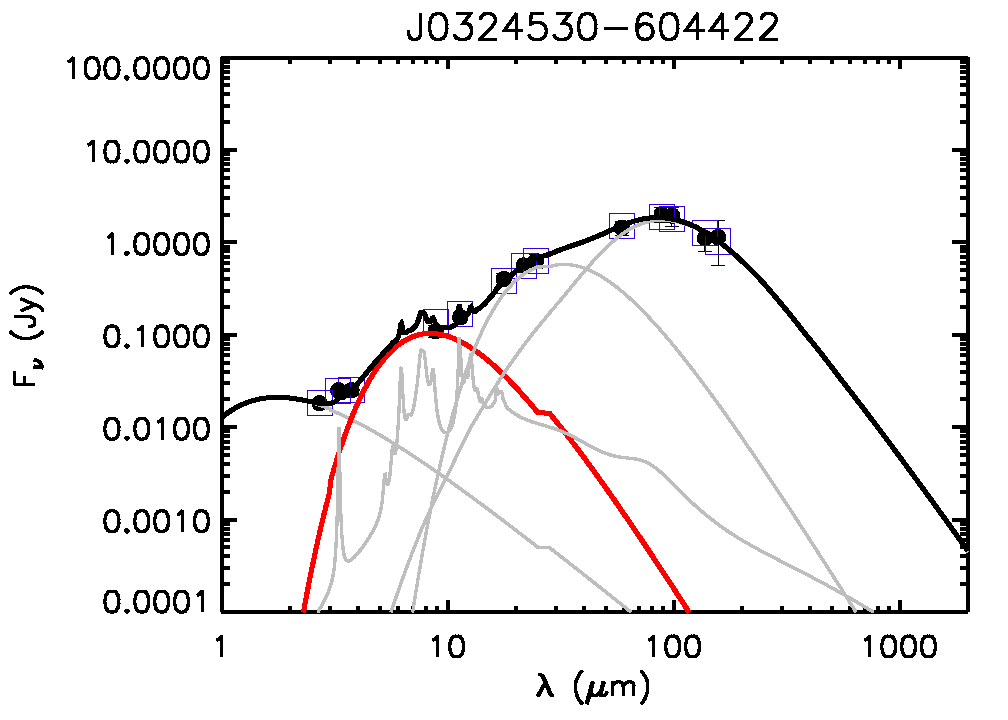}
    \includegraphics[width=5.5cm]{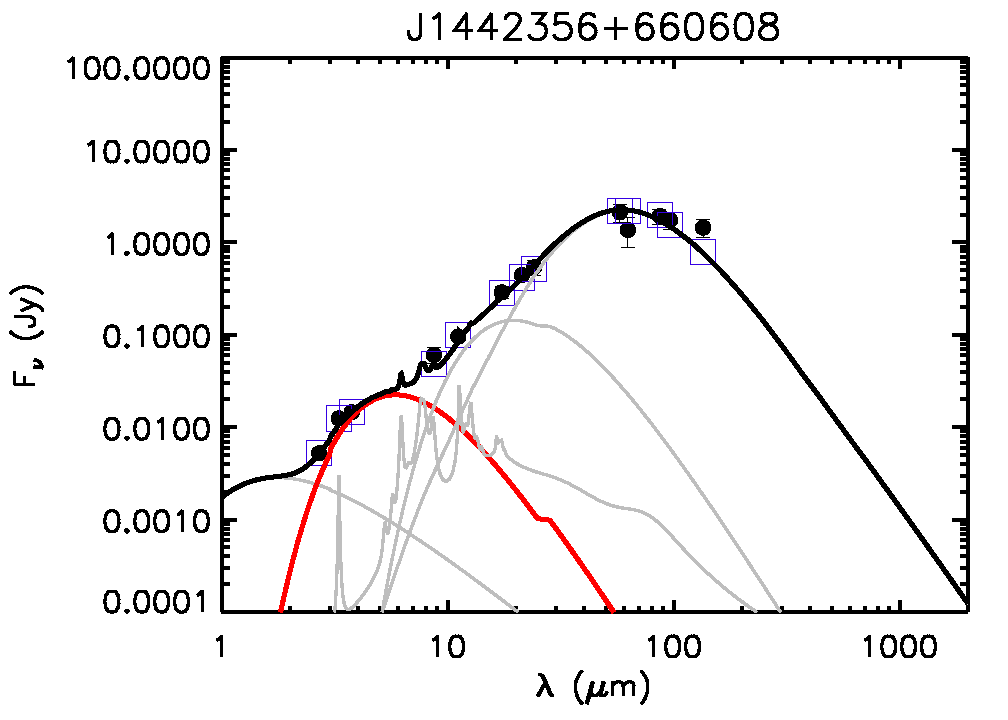} 
    \includegraphics[width=5.5cm]{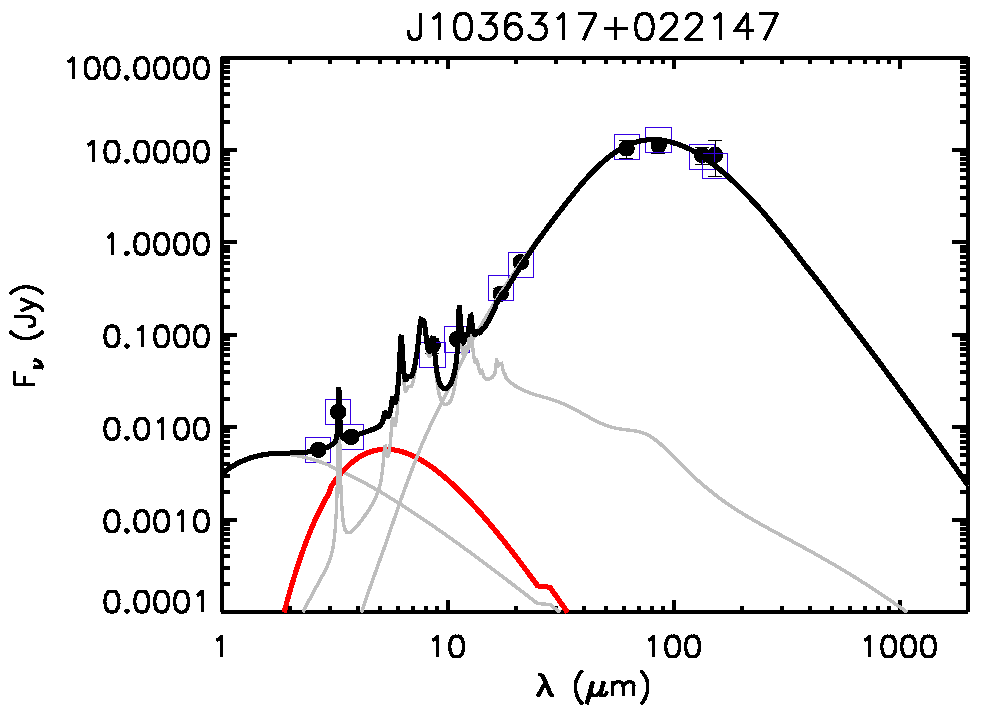} 
    \includegraphics[width=5.5cm]{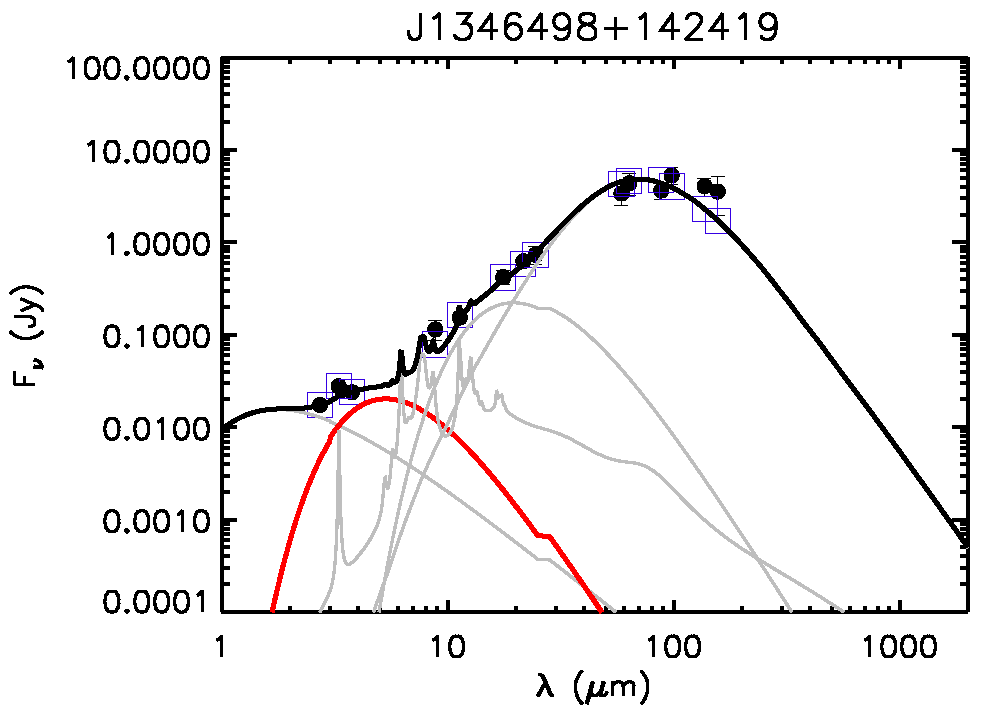}
    \includegraphics[width=5.5cm]{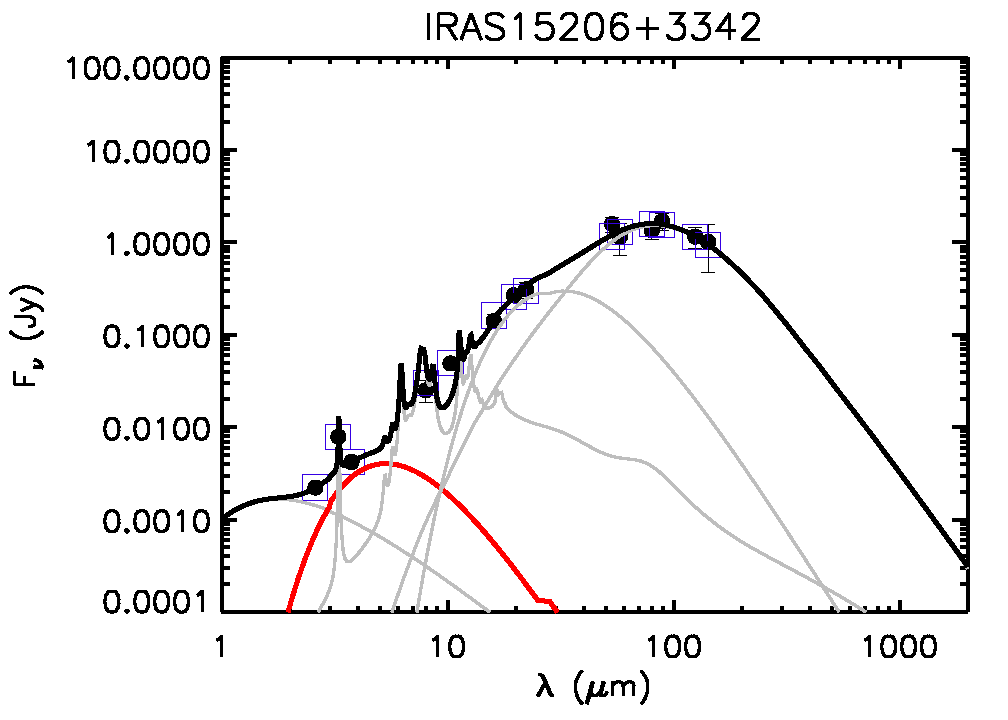}

   \includegraphics[width=5.5cm]{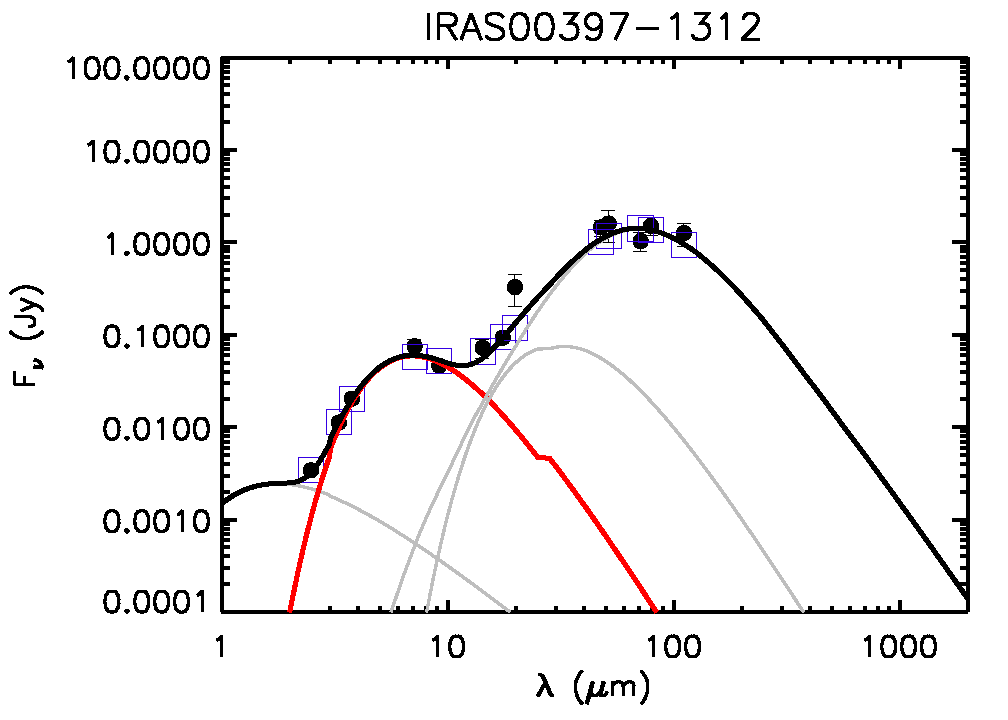} 
    \includegraphics[width=5.5cm]{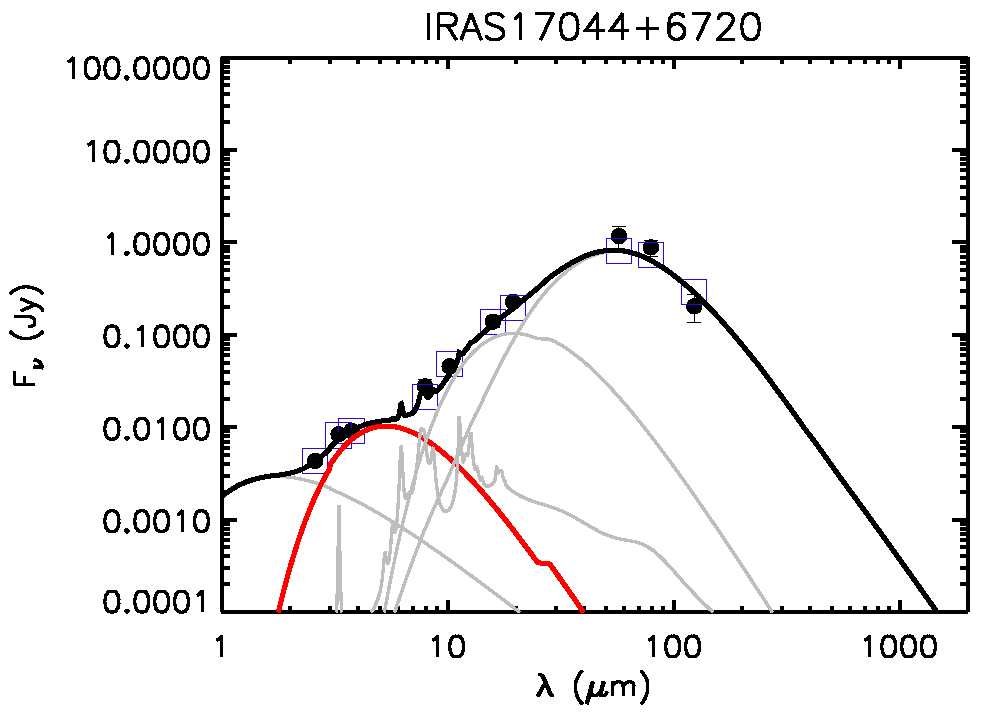} 
    \includegraphics[width=5.5cm]{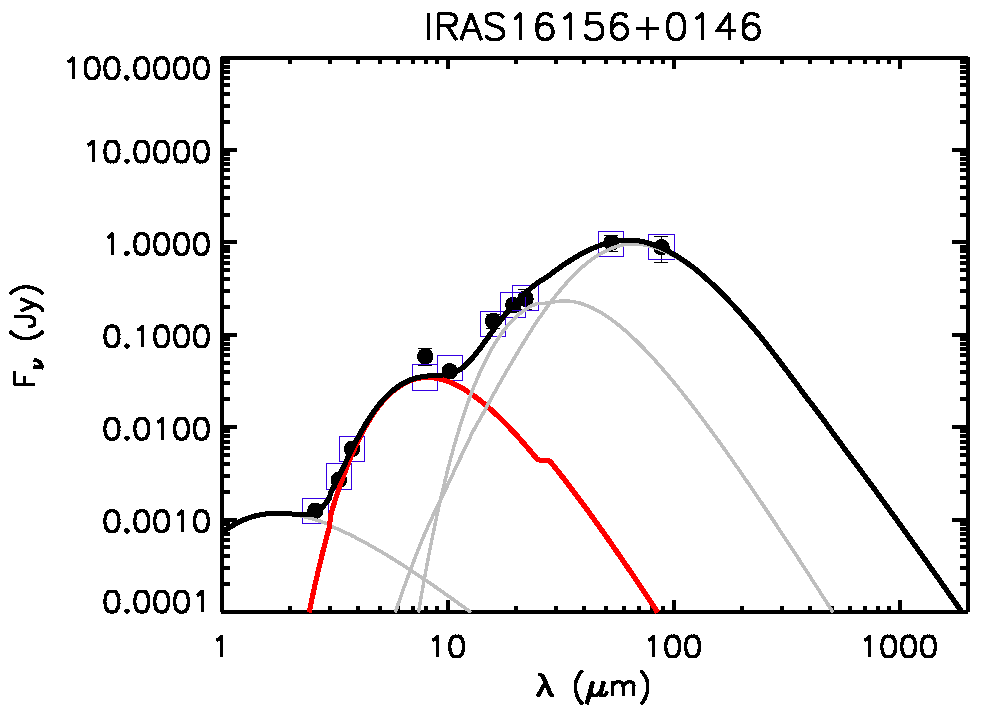}
    \includegraphics[width=5.5cm]{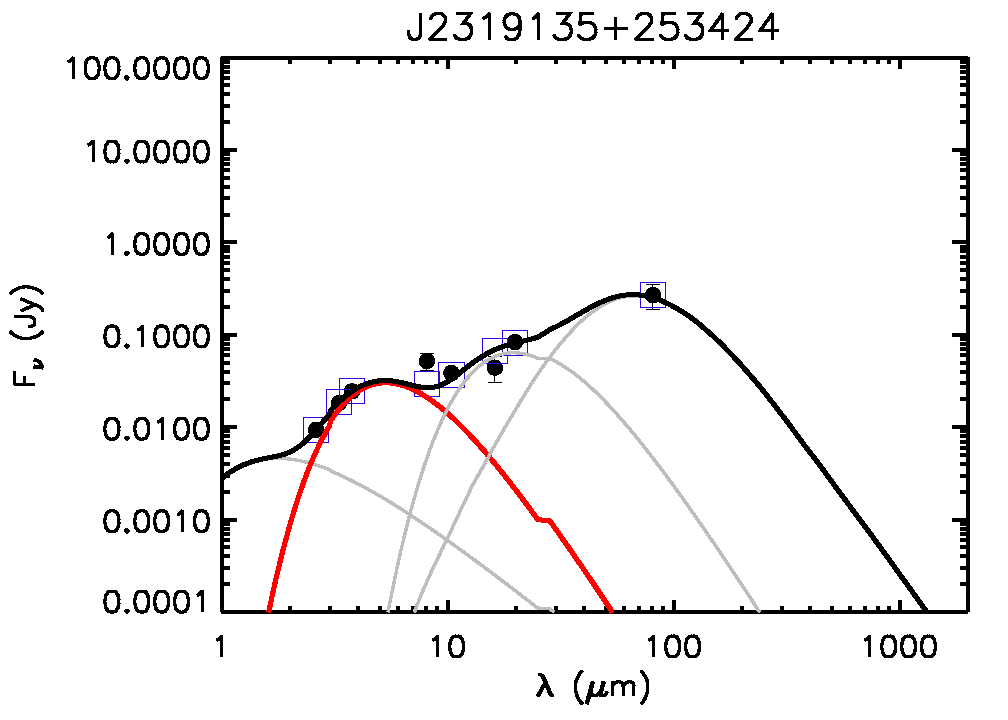}
    \includegraphics[width=5.5cm]{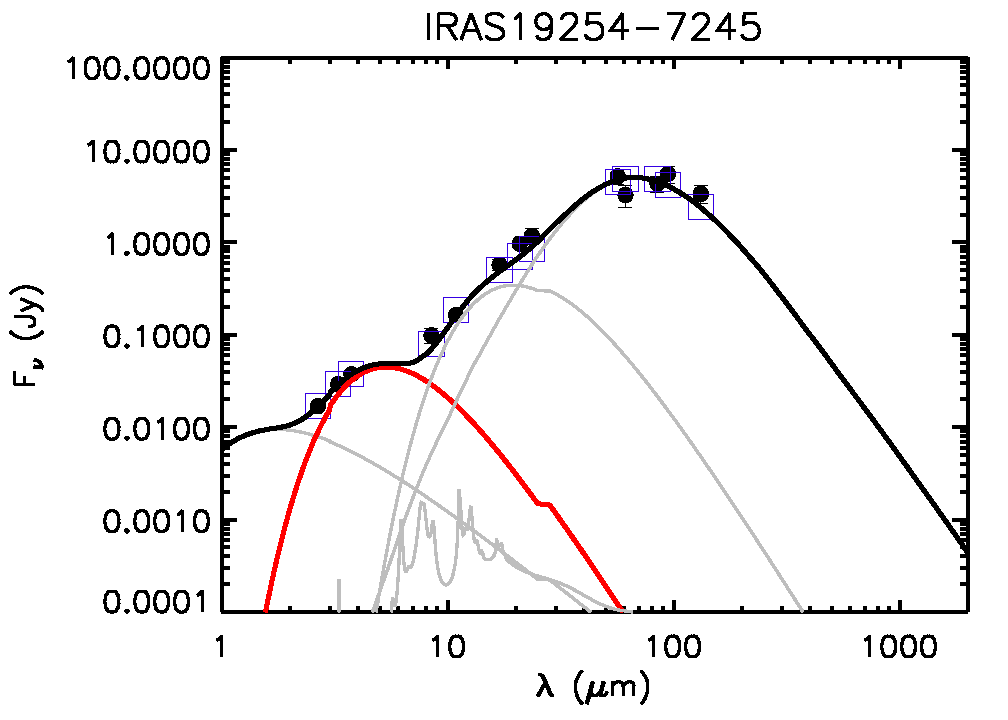} 
    \includegraphics[width=5.5cm]{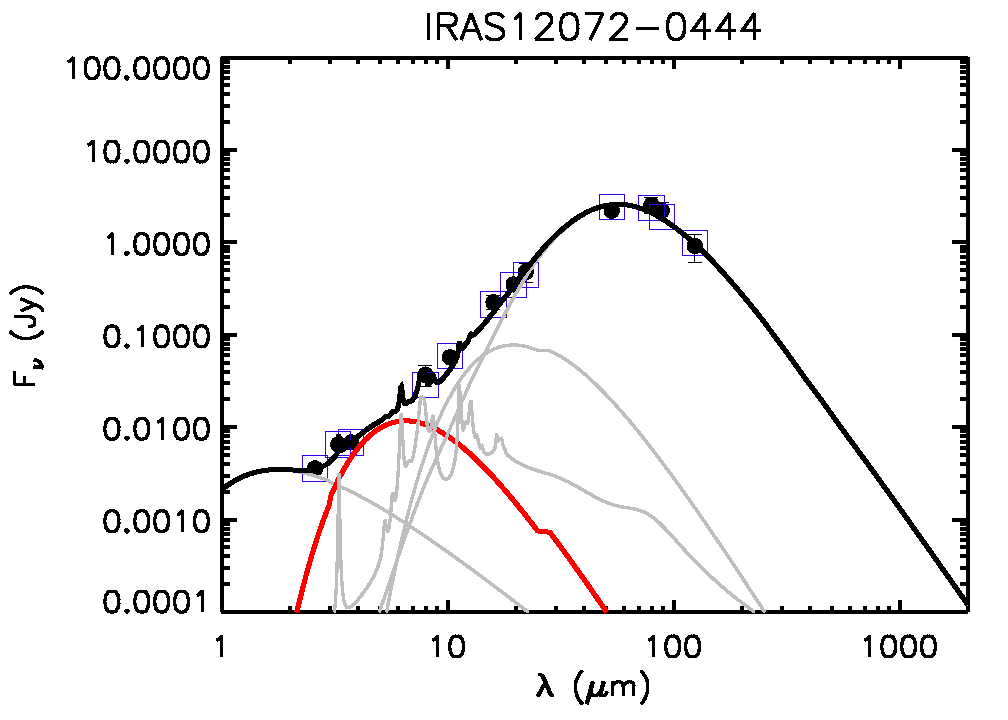}
    \caption{Examples of the SED fitting results for the same galaxies as in Fig. \ref{fig:specfit_sample}.
             A black line shows the total model spectrum,
              while a red line corresponds to the continuum emitted by hot dust. 
             Grey lines show the stellar continuum, 
             neutral and ionized PAHs, the continua emitted by warm dust components, 
             and the continua emitted by SamC, LamC, and aSil, respectively, from short to long wavelengths. 
             Black circles and blue squares represent the fluxes observed in the respective bands 
              and predicted by the model SED, respectively.}
  \label{fig:sed_sample}
  \end{figure*}

  \begin{figure*}[h]
    \centering
    \includegraphics[width=6cm]{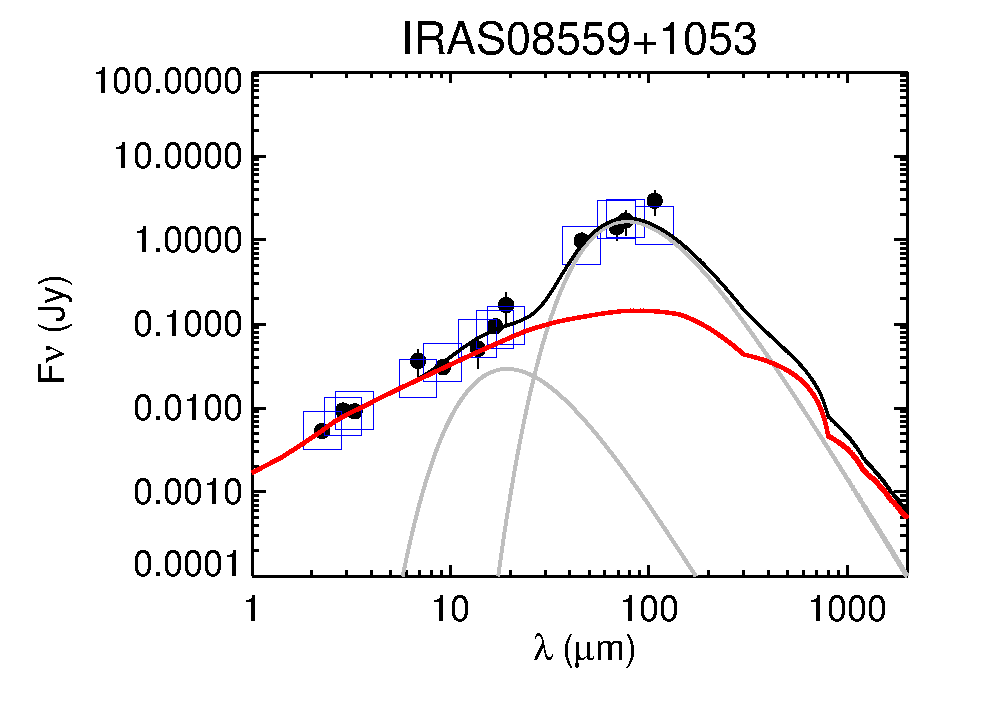}  
    \includegraphics[width=6cm]{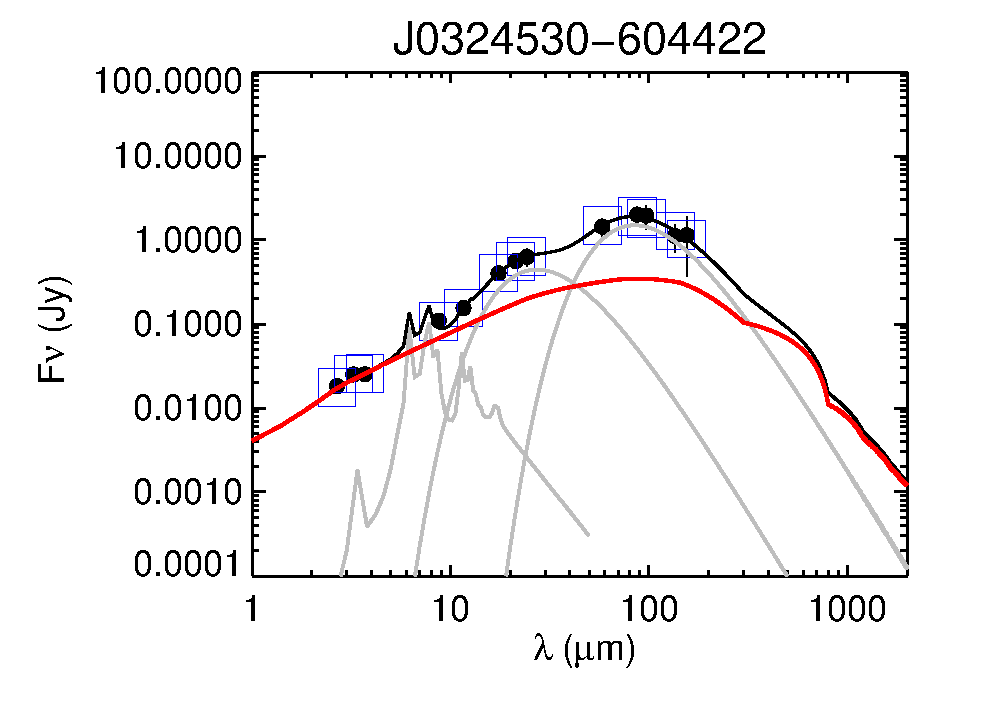}
    \includegraphics[width=6cm]{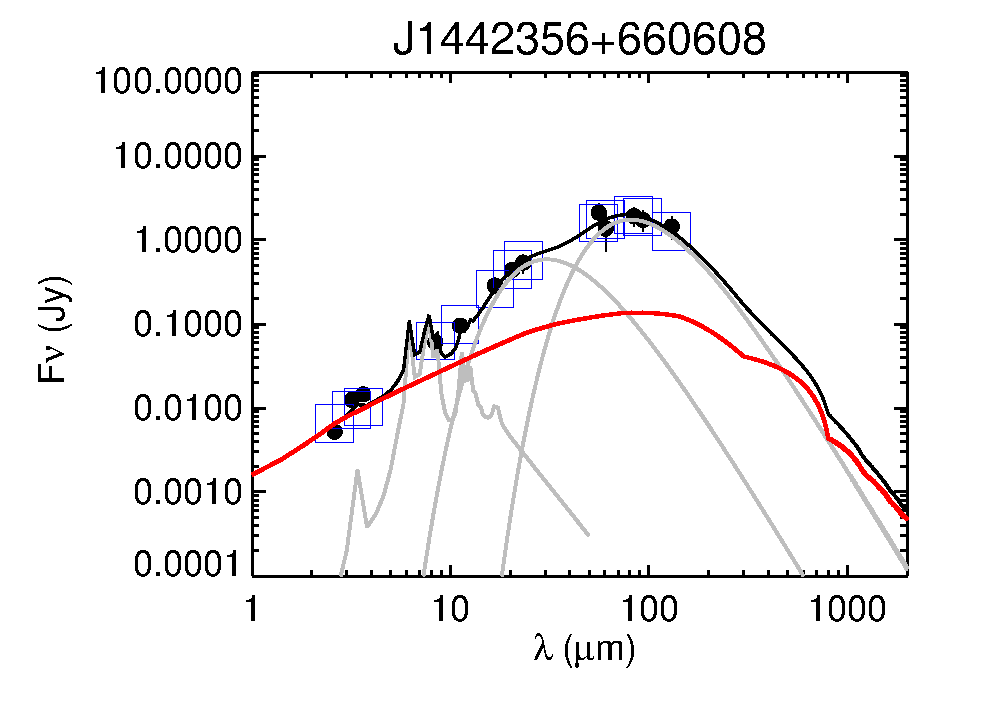} 
    \includegraphics[width=6cm]{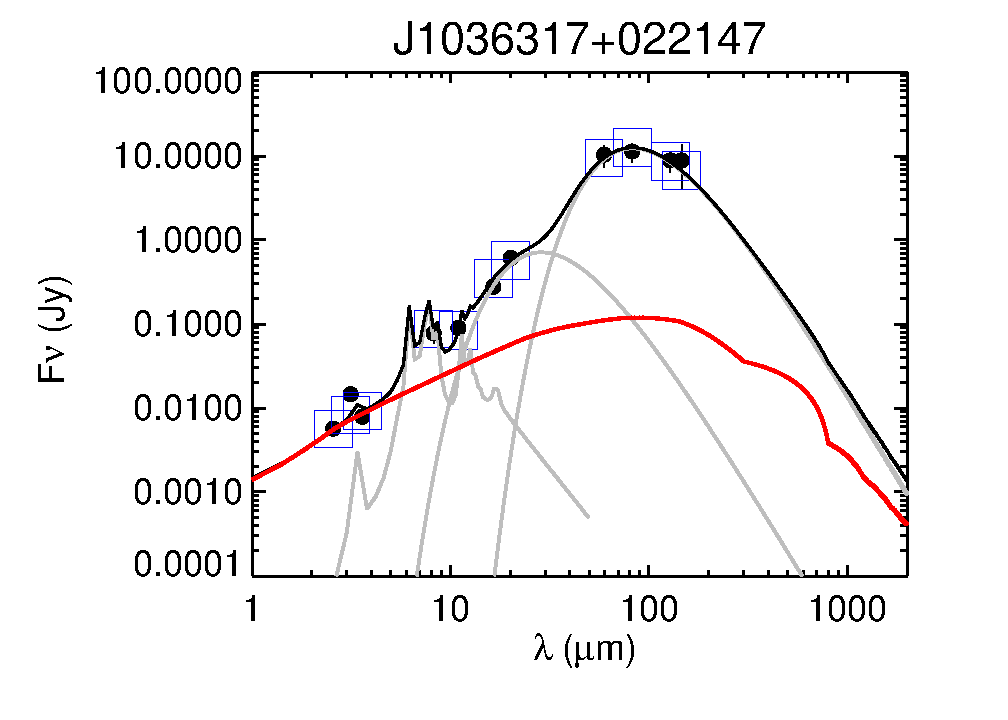} 
    \includegraphics[width=6cm]{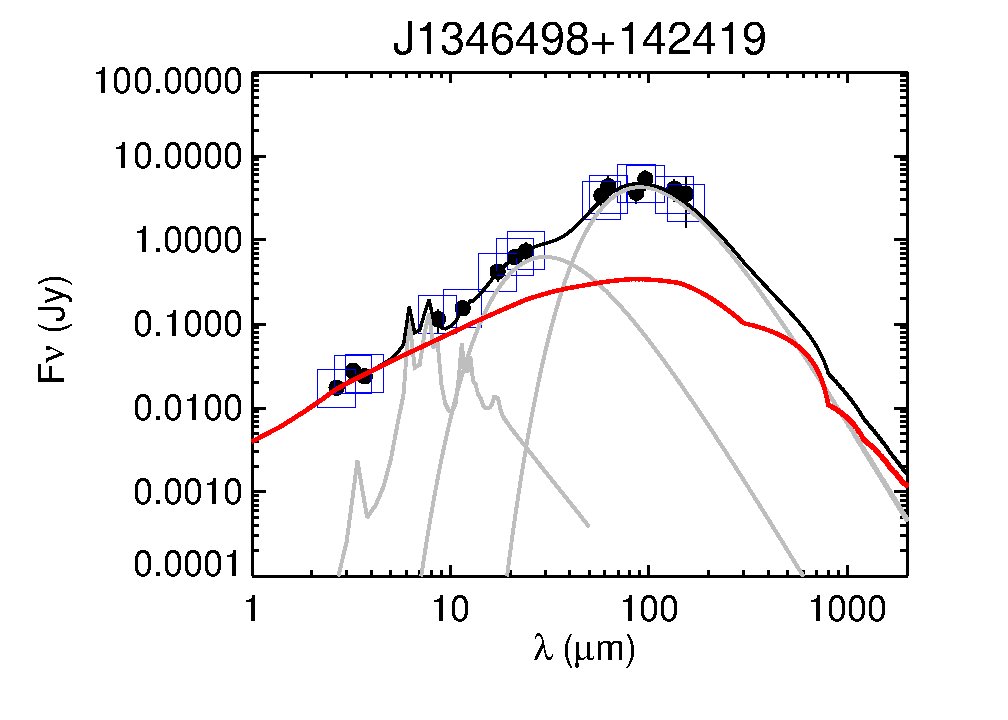}
    \includegraphics[width=6cm]{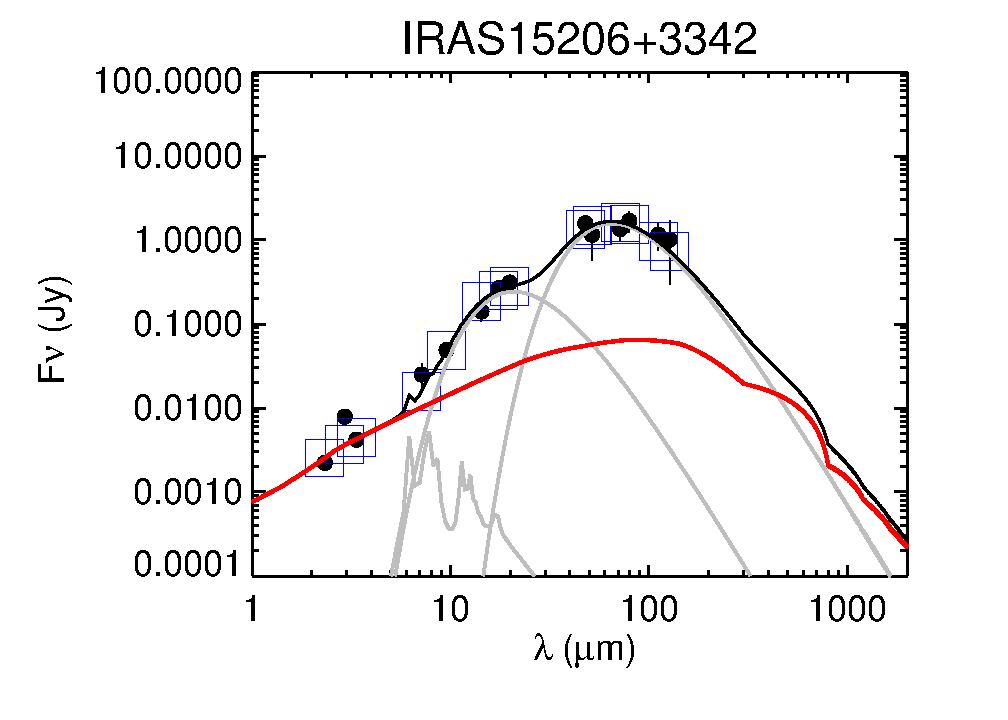}

   \includegraphics[width=6cm]{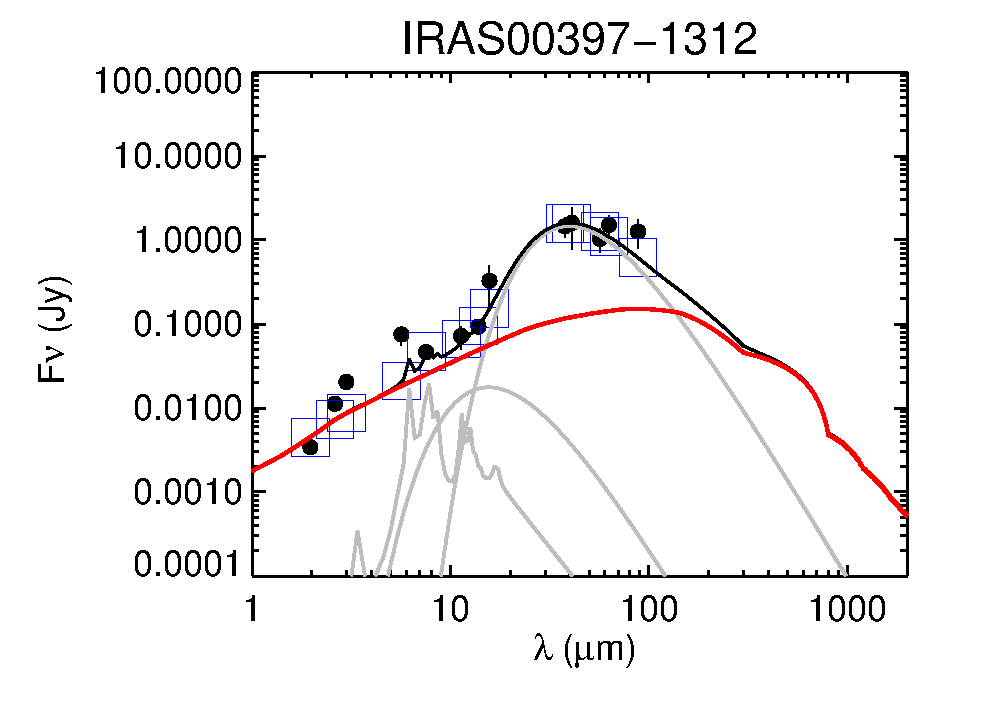} 
    \includegraphics[width=6cm]{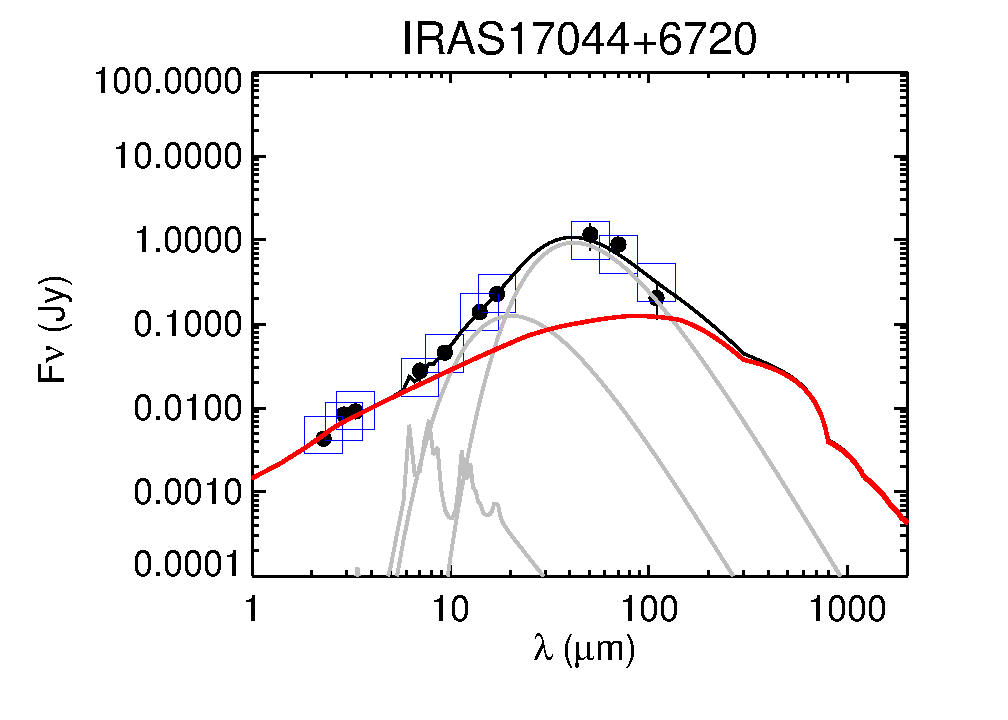} 
    \includegraphics[width=6cm]{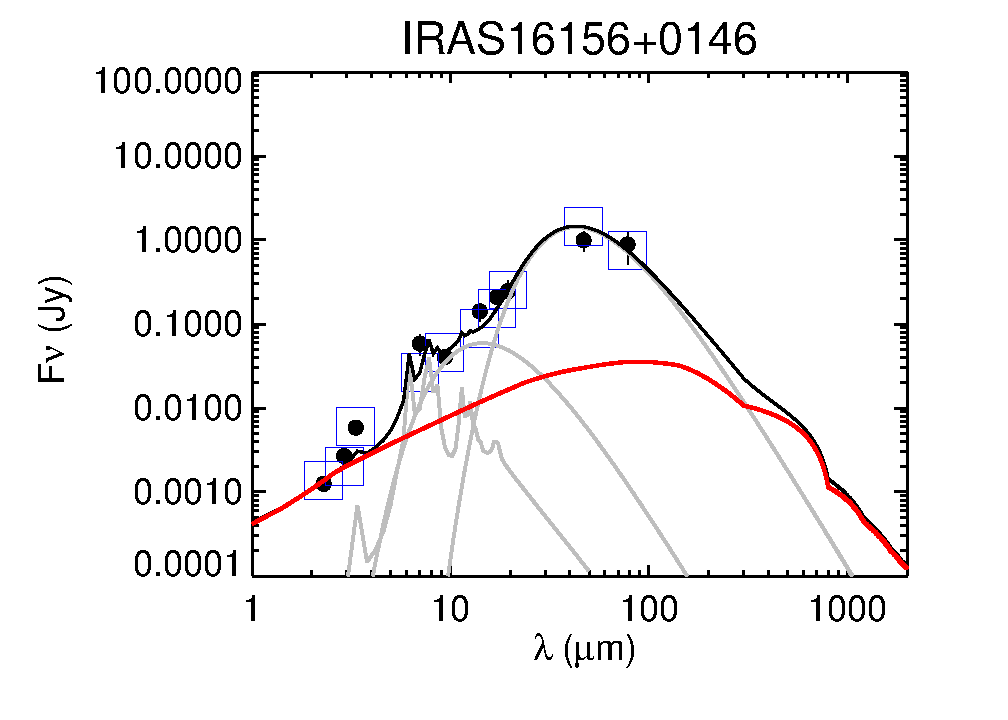}
    \includegraphics[width=6cm]{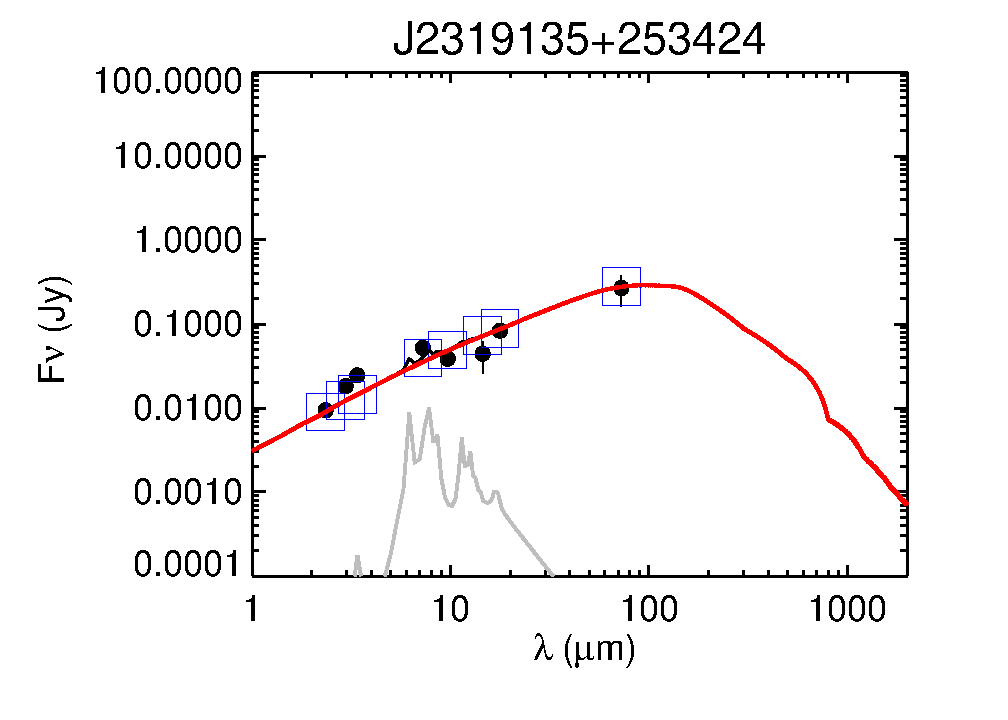}
    \includegraphics[width=6cm]{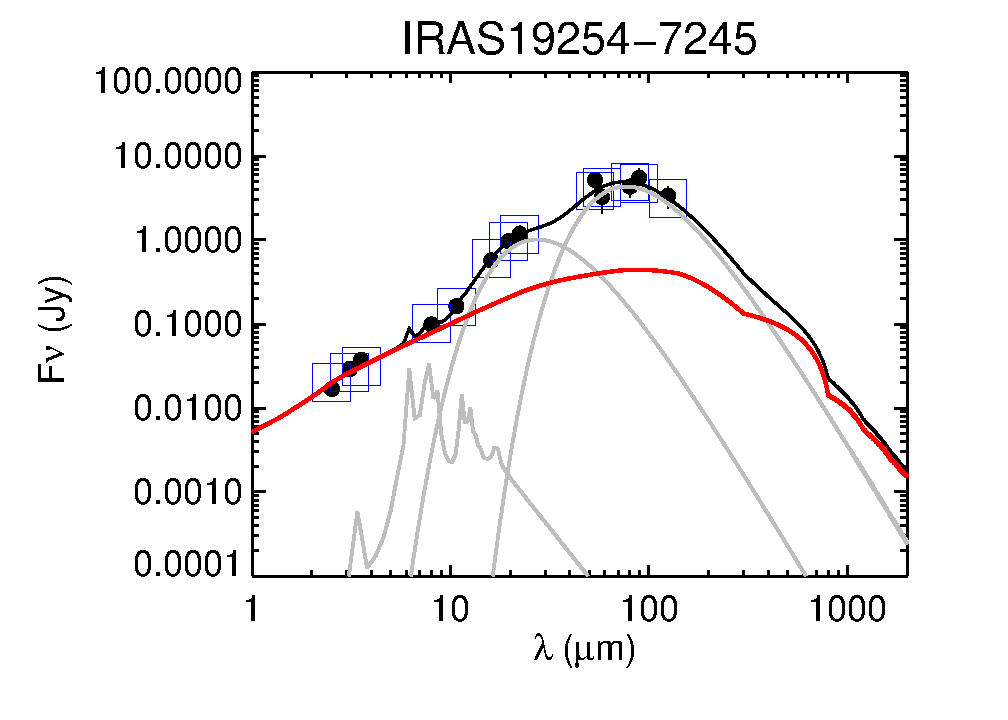} 
    \includegraphics[width=6cm]{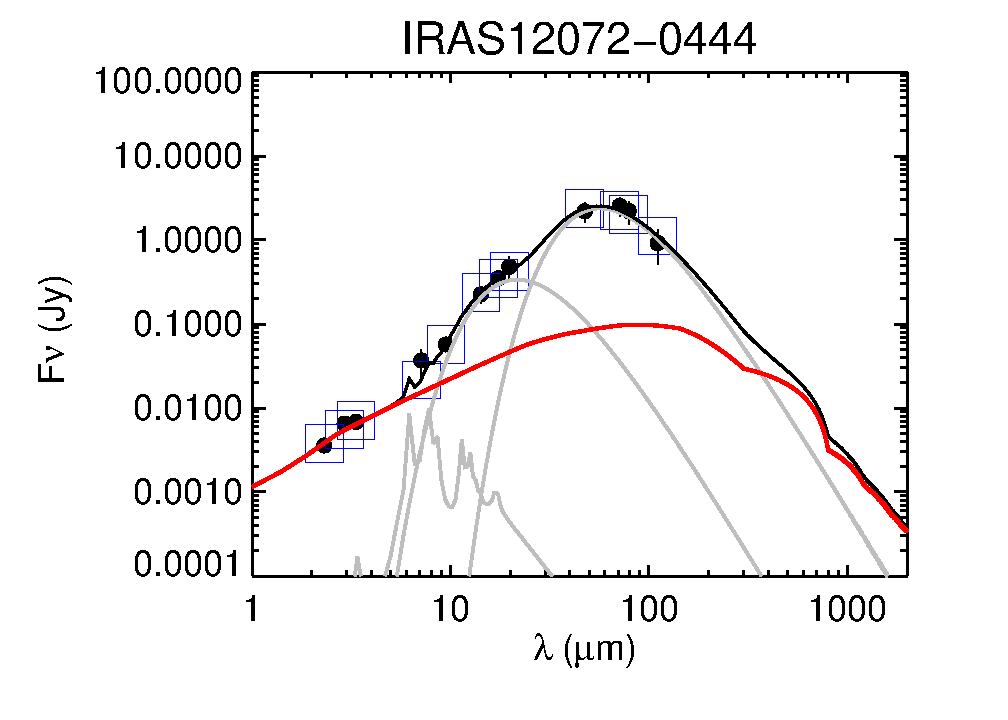}
    \caption{
            Same as Fig. \ref{fig:sed_sample} but the SEDs were fit with a model containing AGN template spectra (see text for details).}
  \label{fig:sed_sample_AGN}
  \end{figure*}
 \begin{table*}
  \caption{Luminosity class and type of our sample galaxies.}
  \begin{tabular}{ccl}
  \hline
  \hline
  Luminosity class & Type & Object name \\
  \hline
  ULIRG & AGN & IRAS 05189-2524, IRAS 23060+0505, IRAS 15130-1958, J 0043041-545826\tablefootmark{(a)},   \\
        &     & J 0047001-174705, J 0356199-625139, J 1040435+593409, J 2141102-264550,  \\
        &     & IRAS 10485-1447, IRAS 00397-1312, IRAS 11223-1244, PKS 1345+12,   \\
        &     & Mrk 231, IRAS 17044+6720, IRAS 13305-1739, \\
        &     & IRAS 23498+2423, IRAS 08572+3915, J 2030417-753243, IRAS 17179+5444\\
        & Composite & IRAS 14202+2615, IRAS 12032+1707, IRAS 14485-2434, IRAS 08559+1053,\\
        &     & IRAS 23389+0300, IRAS 15206+3342, J 1036317+022147, IRAS 04103-2838, \\
        &     & IRAS 11095-0238, IRAS 12359-0725, IRAS 01004-2237, IRAS 12018+1941, \\
        &     & IRAS 12072-0444, IRAS 15001+1433, IRAS 20100-4156, IRAS 01298-0744, \\
        &     & IRAS 14394+5332, IRAS 06035-7102, IRAS 10378+1108, IRAS 16090-0139,\\
        &     & IRAS 22491-1808\\
  \hline
  LIRG  & AGN & NGC 7674, J 0018047-544909, J 0040462-791423, J 0053562-703804, J 0054374-083407,  \\
        &     & IC 5298, J 0522580-362731, J 1041189+574511, J 1050348+801149, J 1353157+634545, \\
        &     & J 2034041-814233, J 2204191+033351, J 2217122+141422, J 2223495-020612, \\
        &     & J 2245112-603357, J 2248045-172829, J 1442356+660608, J 0842153+402532, \\
        &     & J 1407004+282714, J 0109450-033233, J 1628040+514631, J 2319135+253424, \\
        &     & J 1230499+011521, MCG-03-34-064, IRAS 01166-0844, IRAS 19254-7245,  \\
        &     & J 1135491+565710, J 0021533-791007, J 0426007-571203, IRAS 16156+0146, \\
        &     & IRAS 23233+2817 \\
        & Composite & IRAS 20551-4250, IRAS 15250+3609, J 0128477-542131, J 0137067-090853, \\
        &     & J 0334483-561515, IRAS 10173+0828 \\
  \hline
  IRG, sub-IRG & AGN & J 1346498+142419, J 1637082+684253, J 0324530-604422, NGC 4418, J 0023139-535823, \\
               &     &J 1535476+732702,  M 64, M 66, M 94, NGC 383, NGC 7213, NGC 7590 \\
               & Composite & J 0235218-651539\\
  \hline
  \label{tab_sample}
  \end{tabular}\\
  \tablefoot{
    \tablefoottext{a}{The object names starting with J are in the AKARI/IRC all-sky survey point source catalog \citep{Ishihara2010}.}}
  \end{table*} 

\subsection{Near-infrared spectral fitting\label{sect_spec_fit}}
In order to investigate the properties of aromatic and aliphatic hydrocarbons in AGNs, 
 we performed model fitting to the AKARI/IRC 2.5--5.0 $\mu$m spectra,
 assuming the full-screen geometry of obscuration by dust.
In this study, depending on the situation, we used one of the following fitting functions, $F_{\nu}$,:
\begin{equation}
  \centering
 F_{\nu}=e^{-(\tau_{\rm{H_{2}O\ ice}}+\tau_{\rm{CO_{2}\ ice}}+\tau_{\rm{CO\ gas}})}(F_{\rm{cont}}+F_{\rm{aromatic}}+F_{\rm{aliphatic}}+F_{\rm{recomb}})
\label{eq:Fv_em}
\end{equation}
in the case that the aliphatic features are detected in the emission, or
\begin{equation}
  \centering
 F_{\nu}=e^{-(\tau_{\rm{H_{2}O\ ice}}+\tau_{\rm{CO_{2}\ ice}}+\tau_{\rm{CO\ gas}}+\tau_{\rm{aliphatic}})}(F_{\rm{cont}}+F_{\rm{aromatic}}+F_{\rm{recomb}}),
 \label{eq:Fv_ab}
\end{equation}

in the case that the a-C:H absorption features are detected. In the fitting functions, 
$\tau_{\rm{H_{2}O\ ice}}$, $\tau_{\rm{CO_{2}\ ice}}$, and $\tau_{\rm{CO\ gas}}$ 
 are the optical depths for $\rm{H_{2}O\ ice,\ CO_{2}\ ice}$, and CO gas. 
When it was not clear whether the aliphatic features are seen in the emission or absorption by visual inspection,
 we attempted to fit the spectra with each function and chose the one with the better fit.
 The function $F_{\rm{cont}}$ is the flux density representing the stellar and thermal dust continua.
The functions $F_{\rm{aromatic}}$, $F_{\rm{aliphatic}}$, and $F_{\rm{recomb}}$ are the flux densities 
 of the aromatic and aliphatic hydrocarbon emission features and the hydrogen recombination lines, respectively, 
 while $\tau_{\rm{aliphatic}}$ is the optical depth of the a-C:H absorption features.

For the AGN-SF-composite and AGN-dominated galaxies, the global shape of the near-IR continuum varies,
 depending on the fractional contribution of the star formation activity in a host galaxy. 
In order to fit the continuum, we adopted the following function: 
 \begin{equation}
  F_{\rm{cont}}= A\lambda^{\Gamma} + C\frac{B_{\nu}(\lambda,\,T)}{\lambda^{2}},
 \end{equation}
 where $A$ and $C$ are the amplitudes of the power-law and the modified blackbody component, respectively,
and $\Gamma$ is the power-law index.
We added the modified blackbody component to reproduce the emission of the hot dust heated by AGN activity,
 whereas \cite{Kondo2024} used a power-law function alone for the continuum to fit the near-IR spectra of SFGs.
In the case that the fitting result is rejected by the chi-square test,
 we instead adopted a cubic function to fit the continuum.
In fitting the 3.3 $\mu$m aromatic hydrocarbon feature and the 3.4--3.6 $\mu$m aliphatic hydrocarbon features,
 we used the same method as described in \cite{Kondo2024}.
For the aromatic hydrocarbon feature, a Drude profile \citep{Li2001} was adopted:
\begin{equation}
  F_{\rm{aromatic}}=\frac{ab^2}{(\lambda/\lambda_{\rm{r}}-\lambda_{\rm{r}}/\lambda)^2+b^2},
\end{equation}
 where $\lambda_{\rm{r}}$ is the central wavelength fixed at $\lambda_{\rm{r}}=3.3\ \mu$m in the rest frame,
 while $a$ and $b$ are the peak flux density at $\lambda_{\rm{r}}$ and the full width at half maximum divided by $\lambda_{\rm{r}}$, respectively.
\cite{Kondo2024} characterized the 3.4--3.6 $\mu$m aliphatic hydrocarbon features by four representative discrete sub-components
 at the central wavelengths of 3.41, 3.46, 3.51, and 3.56 $\mu$m, referring to previous studies (\citealt{Sloan1997}; \citealt{Li2012}),
 where a Lorentzian profile is used for the first sub-component that is broad enough to be resolved with the AKARI/IRC spectral resolution,
 while Gaussian profiles are used for the other three narrow sub-components.
In order to estimate the flux error of the aliphatic emission feature,
 we fixed the relative strengths of the four sub-components at the best-fit values and refit the spectra.
We assumed the same four sub-components to characterize the a-C:H absorption features,
 which facilitated the comparison between the profiles of the emission and absorption features we present in the discussion.
We confirm that the model of \cite{Dartois2007} for the a-C:H absorption features
 does not improve the spectral fitting significantly.
 We used a Lorentzian profile for the absorption features of $\rm{H_2O}$ ice, $\rm{CO_2}$ ice, and CO gas,
 where the central wavelengths are fixed at 3.05, 4.27, and 4.60 $\mu$m, respectively.
Within the wavelength range of 2.5--5.0 $\mu$m, we considered four relatively strong hydrogen recombination lines, 
 Br$\gamma$, Pf$\gamma$, Br$\alpha$, and Pf$\beta$, 
 at the central wavelengths of 2.62, 3.74, 4.05, and 4.63 $\mu$m, respectively,
 where Gaussian profiles were adopted for all four components.
Examples of the near-IR spectral fitting results for the AGN sample are shown in Fig. \ref{fig:specfit_sample}.

\subsection{Infrared spectral energy distribution fitting}
 We conducted IR spectral energy distribution (SED) fitting using the dust model based on DustEM (v1.2; \citealt{Compiegne2011}).
For the fitting, we used near- to far-IR photometric data obtained with AKARI, 
 the Wide-field Infrared Survey Explorer (WISE; \citealt{Wright2010}), and the Infrared Astronomical Satellite (IRAS; \citealt{Neugebauer1984}),
 which comprise the dataset also used in \cite{Kondo2024}.
In addition, in order to better trace the PAH emission and the hot dust component in the near-IR wavelength,
 we binned the AKARI/IRC spectra to three data points and used them as additional photometric values.
The wavelength, the flux density, and the flux density error of each of the photometric values 
 derived by the spectral binning are defined as follows:
\begin{equation}
  \lambda_i = \frac{\sum\limits_{j=(i-1)N/3+1}^{iN/3}\lambda_j}{N/3},\,
  F_i = \frac{\sum\limits_{j=(i-1)N/3+1}^{iN/3}F_j}{N/3},\,
  \sigma_i = \frac{\sqrt{\sum\limits_{j=(i-1)N/3+1}^{iN/3}\sigma_j^2}}{N/3},
\end{equation}
 where $\lambda_i$, $F_i$, and $\sigma_i$ are the wavelength, the flux density, and the flux density error, respectively.
The subscript $i$ denotes the order of the photometric value after the spectral binning ($i=1, 2, 3$), 
 while $N$ is the total number of the data points included in the AKARI/IRC spectrum.

 The DustEM model contains the following five dust components: 
neutral and ionized PAHs (nPAH and iPAH),
small (sub-nanometer) and large (sub-hundred nanometer) amorphous carbon (SamC and LamC), and amorphous silicate (aSil).
In this paper, for the purpose of better reproducing the continua of stars and the dust components heated by AGNs,
 we added a blackbody and two modified blackbody functions to the original DustEM model,
 where the temperatures were fixed at 3000 K for the stellar continuum 
 and 350--550 K and  60--200 K for the hot and warm dust components, respectively,
 while the amplitudes were allowed to vary. 
The fitting parameters of our model are the mass abundance per hydrogen 
 for each dust component assuming the gas column density of 
 $N_{\mathrm{H}} = 1\times10^{20} \mathrm{cm^{-2}}$ ($Y_{\mathrm{nPAH}},\,Y_{\mathrm{iPAH}},\,Y_{\mathrm{SamC}},\,
  Y_{\mathrm{LamC}},$ and $Y_{\mathrm{aSil}}$), 
 the amplitudes of the blackbody function and the two modified blackbody functions, and the interstellar radiation field
 parameter $G_0$, which is a scaling factor of the radiation field intensity 
 integrated between 6--13 eV relative to the solar neighborhood interstellar radiation field. 
As a result of the SED fitting, we derived the total IR luminosity ($L_{\mathrm{IR}}$), 
 where $L_{\mathrm{IR}}$ is the sum of the luminosities of all the dust components 
 except the stellar continuum.
Examples of the IR SED fitting result for the AGNs and composite galaxies are shown in Fig. \ref{fig:sed_sample}.

Furthermore, in order to estimate the AGN activity, we conducted the SED fitting 
  using another model consisting of the emission components from AGN, PAHs, warm dust, and cold dust.
 For the AGN component, we used the template spectra of type-1 and type-2 AGNs from the SWIRE template library \citep{Polletta2007},
  where the contribution fraction of type-1 and type-2 AGNs and the amplitude are allowed to vary.
 We adopted the \cite{Draine2007} model for the PAH emission and allowed only the amplitude of the PAH emission to vary. 
 For the warm and cold dust components, we adopted modified blackbody models with the emissivity power-law index $\beta = 2$,
  where the amplitudes and temperatures are allowed to vary.
 As a result of the fitting, we derived the luminosity of the AGN component ($L_{\mathrm{AGN}}$).

Finally, out of the 102 AGNs in our original sample, 96 and 90 AGNs 
 were accepted for the spectral fitting and the SED fitting, respectively,
 while 90 AGNs were accepted for both fittings,
 the breakdown of which is 62 AGN-dominated and 28 AGN-SF-composite galaxies, as summarized in Table \ref{tab_sample}.
For the remaining 12 AGNs, the results of the spectral fitting or the SED fitting 
 were rejected through a chi-square test with a confidence level of 90$\%$.
 Most of these AGNs show strong absorption features, and
 their SEDs are difficult to fit successfully with the above SED model.
In order to make a fair comparison between the AGN and SFG samples, 
 we also analyzed the data of the SFG sample presented in \cite{Kondo2024}
 using our spectral and SED fitting models, which are optimized for analyzing the AGN spectra.
As for the SED fitting, we used the above-mentioned first model to estimate $L_{\mathrm{IR}}$ for the AGN and SFG samples,
 while we used the second model to estimate $L_{\mathrm{AGN}}$ for the AGN sample.

\section{Results}
\subsection{Aromatic hydrocarbon emission feature in AGNs\label{sect_result_aro}}
\begin{figure}[h]
  \centering
  \includegraphics[width=7cm]{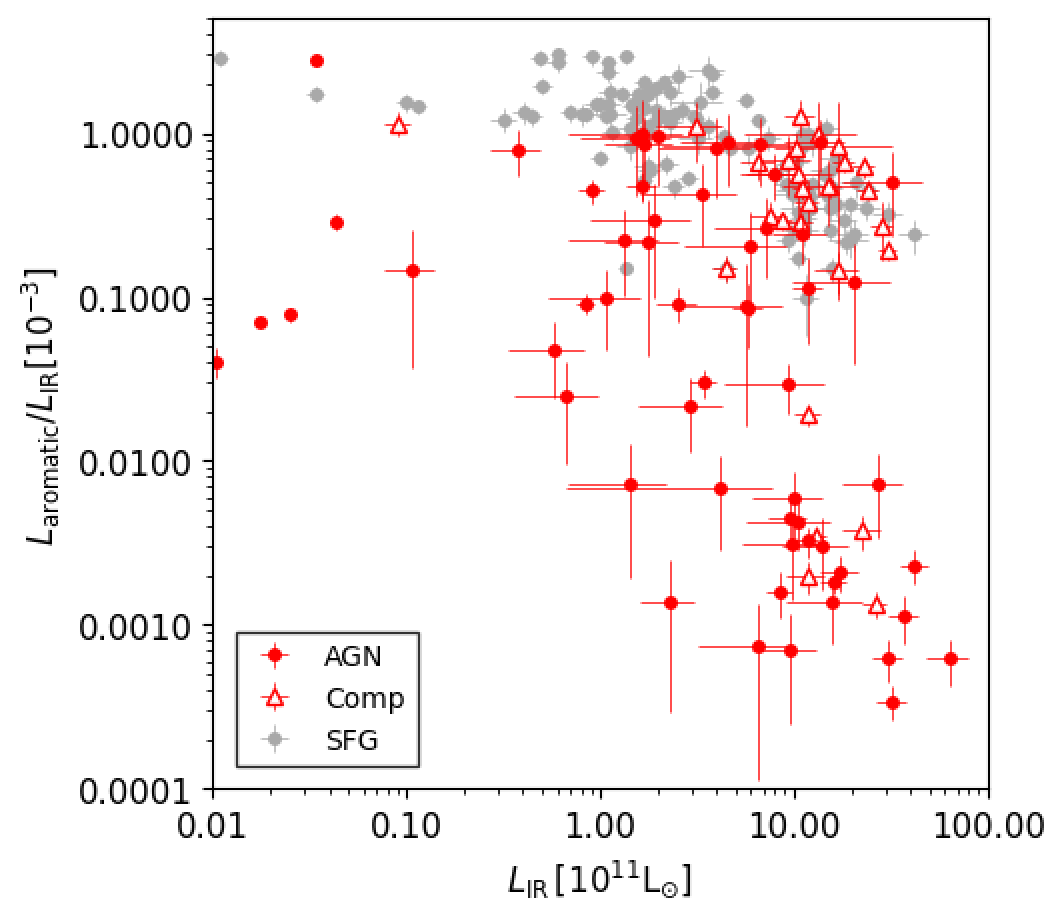} 
  \caption{Scatter plot of $L_{\rm{aromatic}}/L_{\rm{IR}}$ as a function of $L_{\rm{IR}}$
          for the AGN-dominated (red circles) and AGN-SF-composite galaxies (red triangles).
          The gray circles show the relationship for the SFG sample.}
\label{fig:lir_aro}
\end{figure}

Among the 96 AGNs accepted for the spectral fitting, we detected the aromatic feature at 3.3 $\mu$m 
 with a significance of above 3$\sigma$ for 78 AGNs and above 5$\sigma$ for 65 AGNs.
Figure \ref{fig:lir_aro} shows $L_{\mathrm{aromatic}}/L_{\mathrm{IR}}$ as a function of $L_{\mathrm{IR}}$ 
 for all of our sample, which consists of 68 AGN-dominated, 28 AGN-SF-composite, and 122 SFGs.
In comparison with the SFG sample, most of the AGN sample shows a considerably low $L_{\mathrm{aromatic}}/L_{\mathrm{IR}}$
 ($ 10^{-7}\, \lesssim\, L_{\mathrm{aromatic}}/L_{\mathrm{IR}}\, \lesssim\, 10^{-3}$); specifically, it is one to four orders of magnitude lower than $L_{\mathrm{aromatic}}/L_{\mathrm{IR}}$ for the SFG sample.
Regarding the SFG sample, our result shows that the $L_{\mathrm{aromatic}}/L_{\mathrm{IR}}$ significantly decreases 
 with $L_{\mathrm{IR}}$ for LIRGs and ULIRGs ($L_{\mathrm{IR}} > 10^{11}\rm{L_\odot}$),
 which confirms the results in \cite{Yamada2013} and \cite{Kondo2024}, 
 although our models for the spectral and the SED fitting include the hot dust components 
 representative of the AGN contribution in a manner different from models used in previous studies.
The AGN sample also shows a globally decreasing trend with $L_{\mathrm{IR}}$, similar to the SFG sample.

\begin{figure}[h]
  \centering
  \includegraphics[width=7cm]{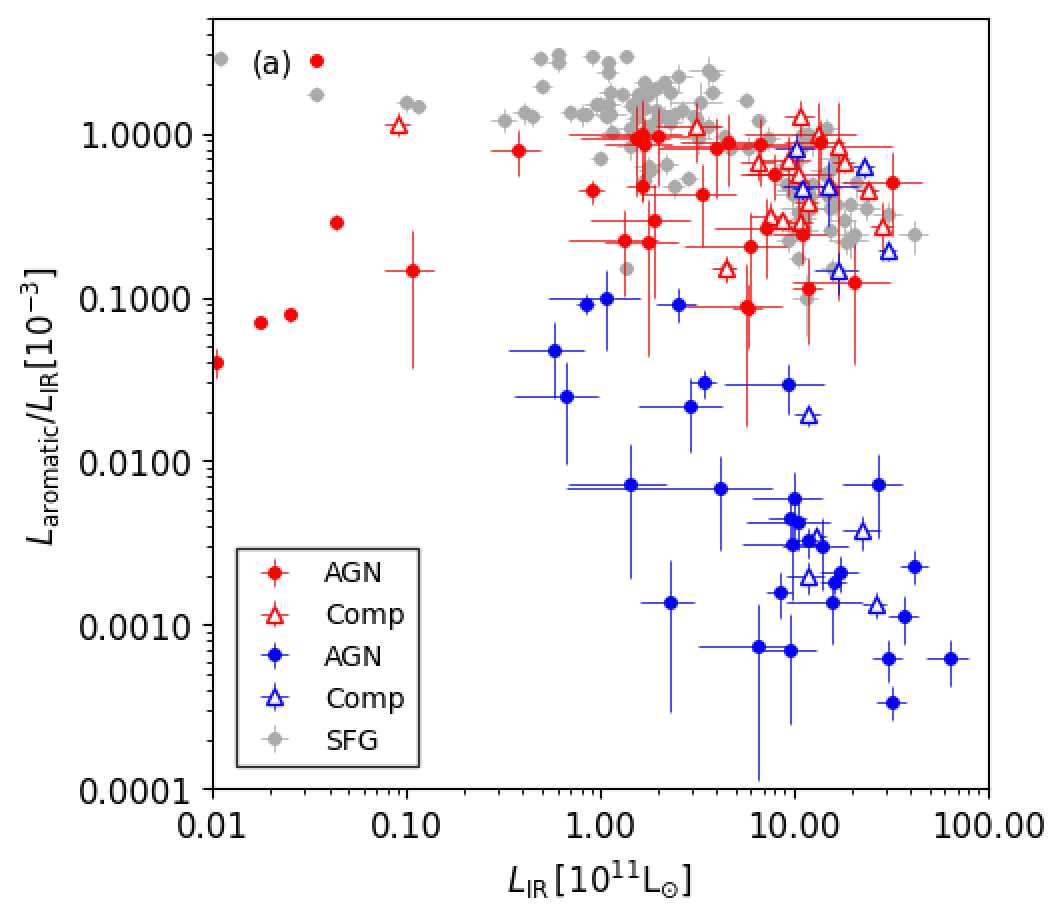} 
  \includegraphics[width=6.7cm]{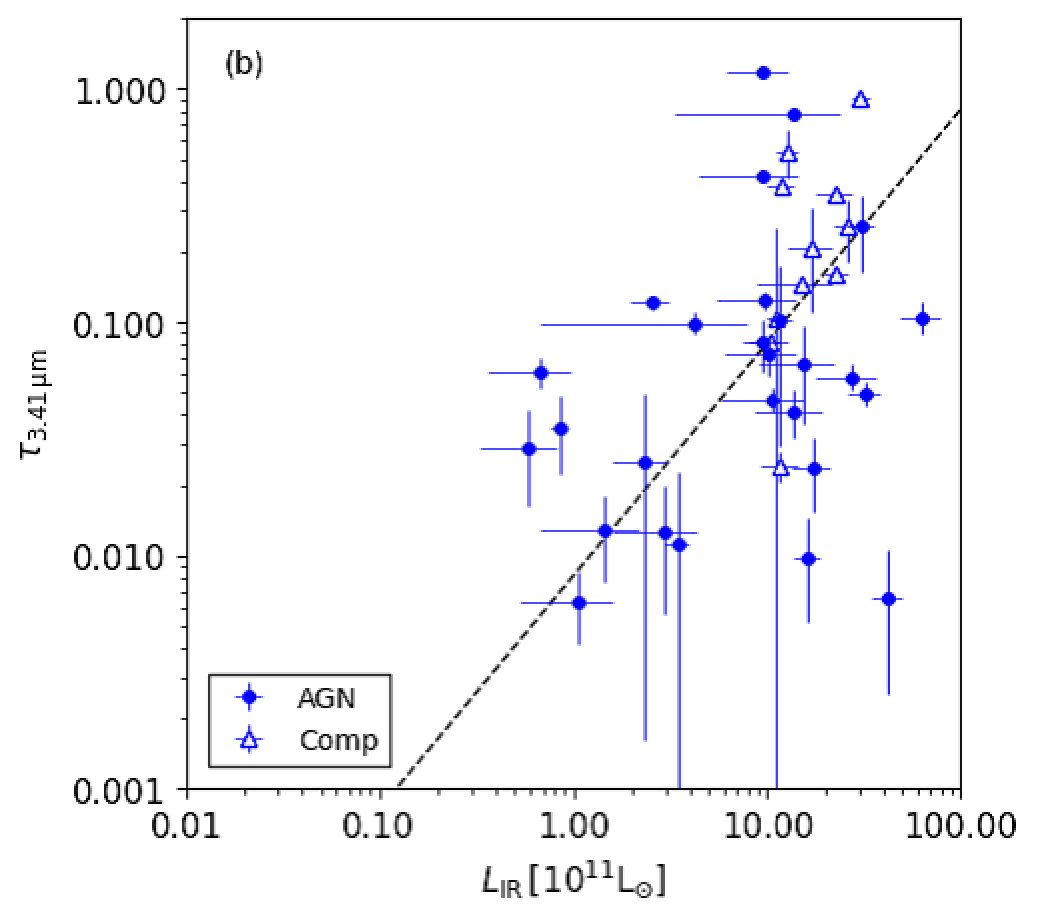}   
  \caption{(a) Same as Fig. \ref{fig:lir_aro} but color coded separately for the aliphatic feature
            detected in the emission (red) and in the absorption (blue).
           (b) Scatter plot of $\tau_{\rm{3.41}}$ as a function of $L_{\rm{IR}}$ for the AGN sample.
           }
\label{fig:lir_aro_ab}
\end{figure}

\begin{figure*}[h]
  \centering
 
  \includegraphics[width=8.2cm]{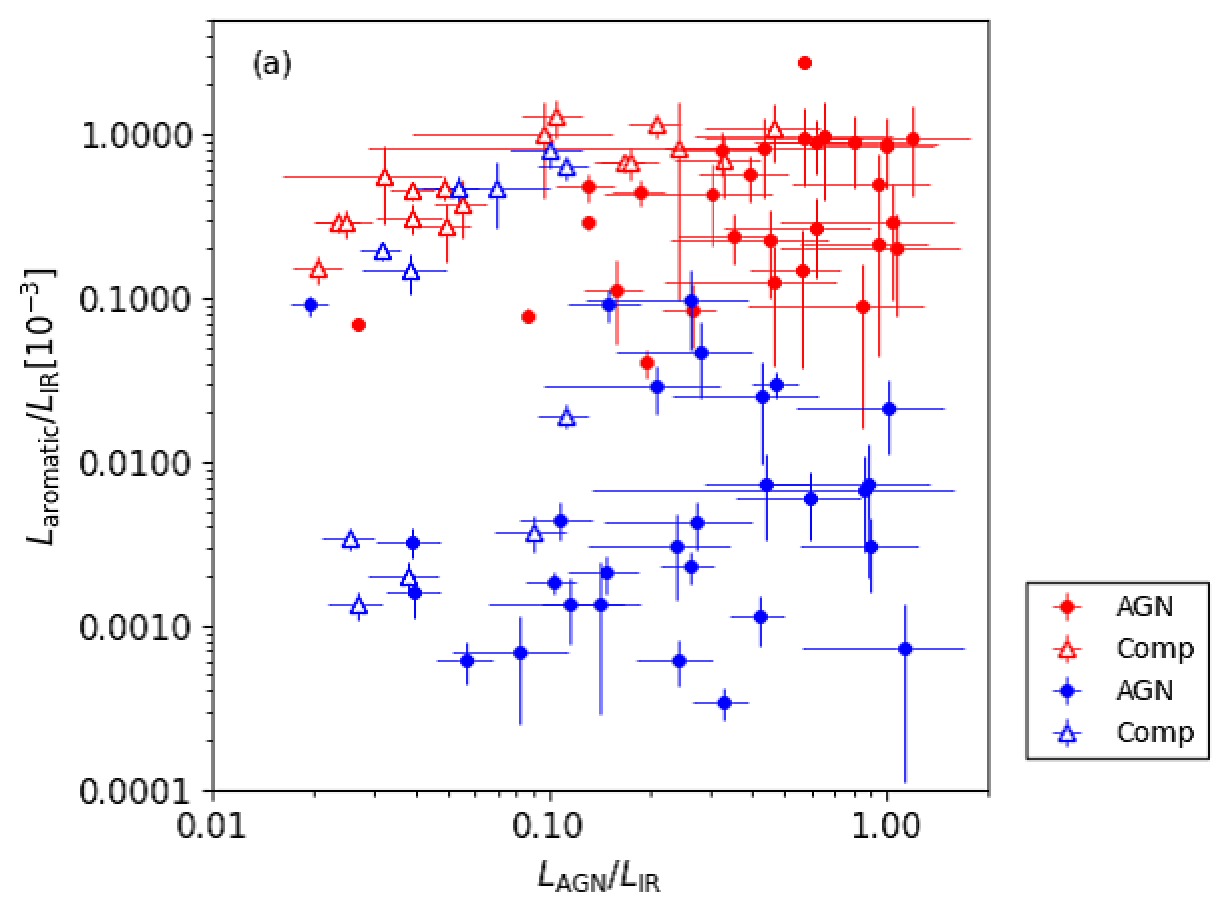}
  \includegraphics[width=7cm]{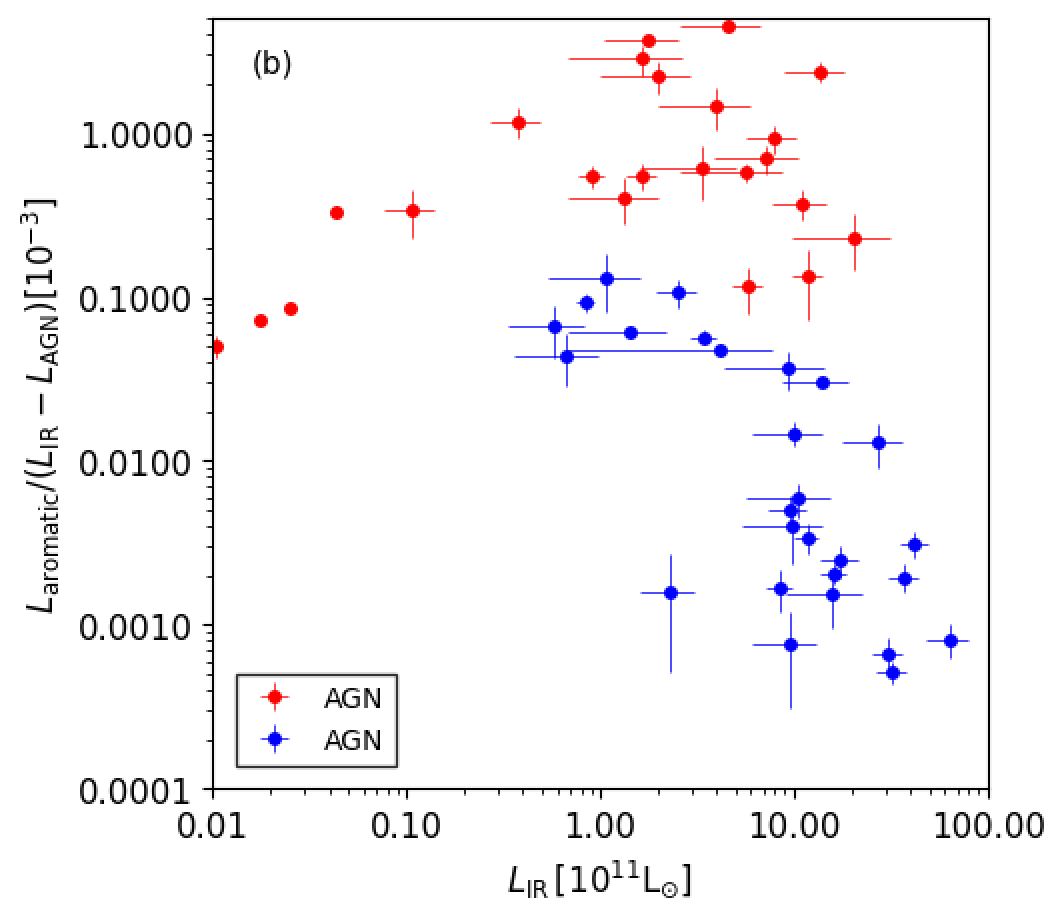}
  \caption{(a) Same as Fig. \ref{fig:lir_aro_ab}a but plotted as a function of $L_{\rm{AGN}}/L_{\rm{IR}}$ for the AGN sample.
           (b) Same as Fig. \ref{fig:lir_aro_ab}a, but $L_{\rm{aromatic}}/L_{\rm{IR}}$ is replaced with $L_{\rm{aromatic}}/(L_{\rm{IR}}-L_{\rm{AGN}})$
  \label{fig:lir_tali1}
           and plotted for the AGN-dominated galaxies.} 
\end{figure*}
In Fig. \ref{fig:lir_aro_ab}a, we plot $L_{\mathrm{aromatic}}/L_{\mathrm{IR}}$  and color code it separately
 for the samples with the aliphatic features detected in the emission (red) and 
 the a-C:H features detected in the absorption (blue).
From the figure, we observed that the $L_{\mathrm{aromatic}}/L_{\mathrm{IR}}$ distributions of
 the aliphatic emission and the a-C:H absorption AGN samples are clearly separated from each other.
The aliphatic emission AGN sample (red) shows a relatively high $L_{\mathrm{aromatic}}/L_{\mathrm{IR}}$, but it is still systematically lower than that for the SFG sample (gray) 
 in the IRG and LIRG luminosity classes.
The $L_{\mathrm{aromatic}}/L_{\mathrm{IR}}$ values of the aliphatic emission AGN sample have no significant dependence on $L_{\mathrm{IR}}$,
 the trend of which is different from that of the SFG sample where $L_{\mathrm{aromatic}}/L_{\mathrm{IR}}$ decreases with $L_{\mathrm{IR}}$.
To evaluate the statistical significance, we applied a Kolmogorov-Smirnov test (KS test)
to confirm that their distributions are different with a significance level of 5$\%$.

On the other hand, most of the a-C:H absorption AGN sample (blue) in Fig. \ref{fig:lir_aro_ab}a show a considerably low $L_{\mathrm{aromatic}}/L_{\mathrm{IR}}$.
Among the a-C:H absorption AGN sample, 
 six galaxies show a $L_{\mathrm{aromatic}}/L_{\mathrm{IR}}$ as high as the SFG sample at around $L_{\mathrm{IR}}=10^{12}\rm{L_\odot}$, and
 all of these galaxies belong to the composite galaxy population.
Except for those galaxies, $L_{\mathrm{aromatic}}/L_{\mathrm{IR}}$ significantly decreases with $L_{\mathrm{IR}}$.
Figure \ref{fig:lir_aro_ab}b shows $\tau_{3.41}$ plotted as a function of $L_{\mathrm{IR}}$ for the a-C:H absorption AGN sample,
 from which we find that $\tau_{3.41}$ significantly increases with $L_{\mathrm{IR}}$.
The comparison of Figs. \ref{fig:lir_aro_ab}a and \ref{fig:lir_aro_ab}b suggests that 
 the decrease in $L_{\mathrm{aromatic}}/L_{\mathrm{IR}}$ for the a-C:H absorption AGN sample in Fig. \ref{fig:lir_aro_ab}a
 may be attributed to the interstellar dust extinction represented by $\tau_{3.41}$.
As seen in Fig. \ref{fig:lir_aro_ab}b, $\tau_{3.41}$ is large for ULIRGs ($L_{\rm{IR}} > 10^{12}\rm{L_\odot}$),
 which are basically dust-rich galaxies, 
 supporting that the absorbing aliphatic hydrocarbons are likely of interstellar dust origins rather than of circumnuclear ones.

In Fig. \ref{fig:lir_tali1}a, we plot $L_{\mathrm{aromatic}}/L_{\mathrm{IR}}$ as a function of $L_{\rm{AGN}}/L_{\rm{IR}}$.
For a galaxy hosting an AGN, $L_{\rm{AGN}}/L_{\rm{IR}}$ corresponds to the fractional luminosity of the dust heated by the AGN,
 and thus it is considered a likely indicator of AGN activity.
As shown in the figure, $L_{\mathrm{aromatic}}/L_{\mathrm{IR}}$ is not clearly correlated with $L_{\rm{AGN}}/L_{\rm{IR}}$
 for either aliphatic emission or a-C:H absorption samples.
For AGNs with relatively low AGN activity ($L_{\rm{AGN}}/L_{\rm{IR}}\,\la\,0.1$), 
 the $L_{\mathrm{aromatic}}/L_{\mathrm{IR}}$ ratios are bifurcated into low and high levels,
 the latter of which mostly belong to the composite galaxy population.
In Fig. \ref{fig:lir_tali1}b, considering a possibility that the decrease in $L_{\mathrm{aromatic}}/L_{\mathrm{IR}}$ for the absorption AGN sample in Fig. \ref{fig:lir_aro_ab}a
   could be explained by a larger contribution of $L_{\rm{AGN}}$ to $L_{\rm{IR}}$,
   we also plot $L_{\mathrm{aromatic}}/(L_{\mathrm{IR}}-L_{\mathrm{AGN}})$ as a function of $L_{\mathrm{IR}}$,
   from which we confirm that the decreasing trend still holds.

Hence, most of the AGN-dominated galaxies with the a-C:H features detected in the absorption, 
 especially those in the ULIRG class ($L_{\rm{IR}} > 10^{12}\rm{L_\odot}$),
 have very low $L_{\mathrm{aromatic}}/L_{\mathrm{IR}}$ values (< $10^{-5}$),
 suggesting that aromatic hydrocarbons exist mainly in the circumnuclear region
 and that their emission is strongly suppressed due to the interstellar dust extinction in the outer regions of the host galaxy.
This is likely to also be the case for the AGN-dominated galaxies with the aliphatic feature detected in the emission.
As seen in Fig. \ref{fig:lir_tali1}a, for AGNs with relatively high activity ($L_{\rm{AGN}}/L_{\rm{IR}}$\, $\gtrsim\, 0.2$),
 the $L_{\mathrm{aromatic}}/L_{\mathrm{IR}}$ values are instead continuously distributed within the range of $10^{-7}$ to $10^{-3}$,
 which may be explained by the difference in the inclination angle of the galactic disk with respect to the line of sight (i.e., near edge-on or face-on configuration). 
On the other hand, some of the AGN-SF-composite galaxies show a relatively high $L_{\mathrm{aromatic}}/L_{\mathrm{IR}}$ 
 whether the aliphatic feature is detected in the emission or the a-C:H feature is detected in the absorption,
 and therefore those galaxies are considered to have aromatic hydrocarbons distributed more widely in the outer interstellar regions of a host galaxy.

\subsection{Aliphatic hydrocarbon emission features in AGNs\label{sect:ali_em}}
   
\begin{figure*}[h]
  \centering
  \includegraphics[width=7cm]{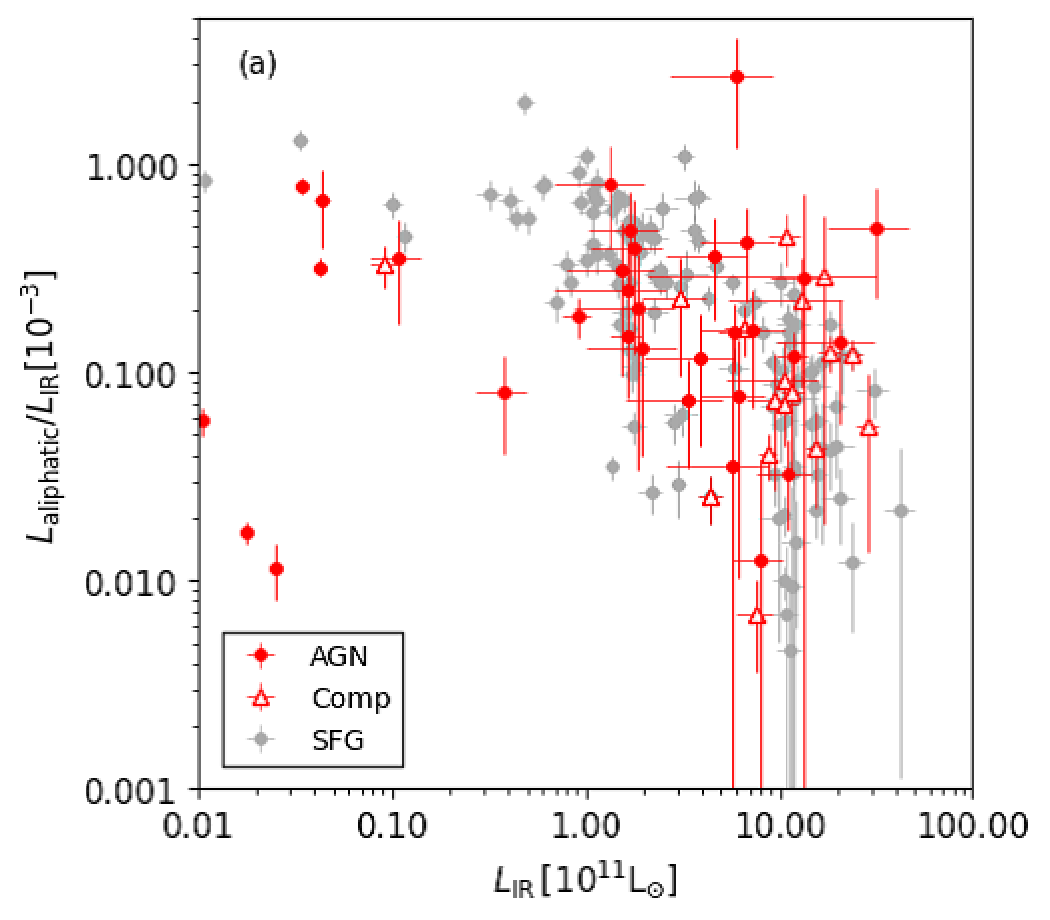} 
  \includegraphics[width=6.7cm]{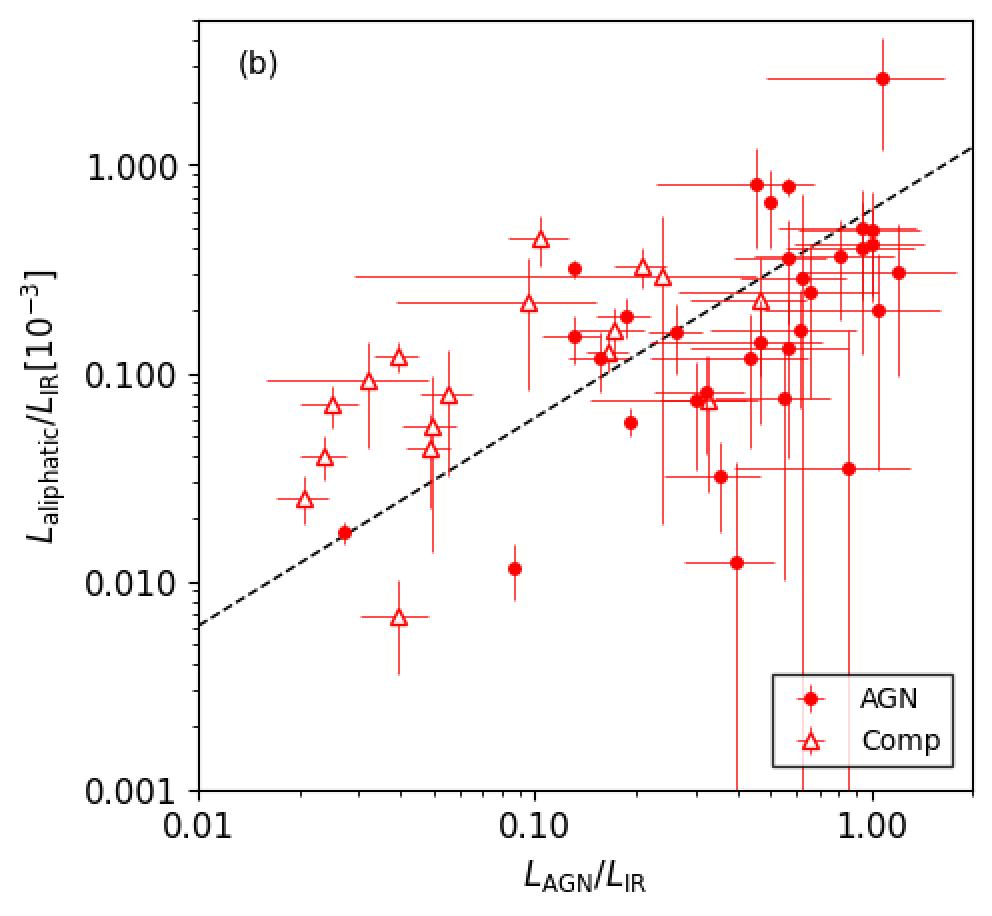}  
  \caption{Scatter plots of $L_{\rm{aliphatic}}/L_{\rm{IR}}$ as functions of (a) $L_{\rm{IR}}$ and (b) $L_{\rm{AGN}}/L_{\rm{IR}}$
          for the AGN-dominated (red circles) and AGN-SF-composite galaxies (red triangles).
          The gray circles in panel (a) show the relationship for the SFG sample.}
\label{fig:lir_ali}
\end{figure*}
We show the result of the spectral fitting on the aliphatic emission features in Fig. \ref{fig:lir_ali}, 
 where $L_{\rm{aliphatic}}/L_{\rm{IR}}$ is plotted as functions of 
  $L_{\rm{IR}}$ and $L_{\rm{AGN}}/L_{\rm{IR}}$ in panels (a) and (b), respectively.
For the SFG sample, we also plot the result of our re-analysis in panel (a),
  which exhibits that $L_{\rm{aliphatic}}/L_{\rm{IR}}$ decreases with $L_{\rm{IR}}$, which is consistent with the result in \cite{Kondo2024}.
As a whole, $L_{\rm{aliphatic}}/L_{\rm{IR}}$ shows similar values between the AGN and SFG samples.
The KS test shows that the distributions of the AGN and SFG samples are not different from each other, with a significance level of 5$\%$.
Unlike the SFG sample, $L_{\rm{aliphatic}}/L_{\rm{IR}}$ of the AGN sample shows no apparent dependence on $L_{\rm{IR}}$ (Fig. \ref{fig:lir_ali}a)
 but an increasing trend with $L_{\rm{AGN}}/L_{\rm{IR}}$ (Fig. \ref{fig:lir_ali}b).
The latter result suggests that the emitting aliphatic hydrocarbons may be of circumnuclear origin and significantly affected by AGN activity,
 while the absorbing aliphatic hydrocarbons are likely of interstellar origin, 
 as mentioned in Section \ref{sect_result_aro}. 
It is notable that the former hydrocarbons are more likely to be produced than destroyed by the AGN activity.

\subsection{Relationships between the aliphatic-to-aromatic ratios and the AGN activity}
\begin{figure}[h]
 \centering
 \includegraphics[width=7cm]{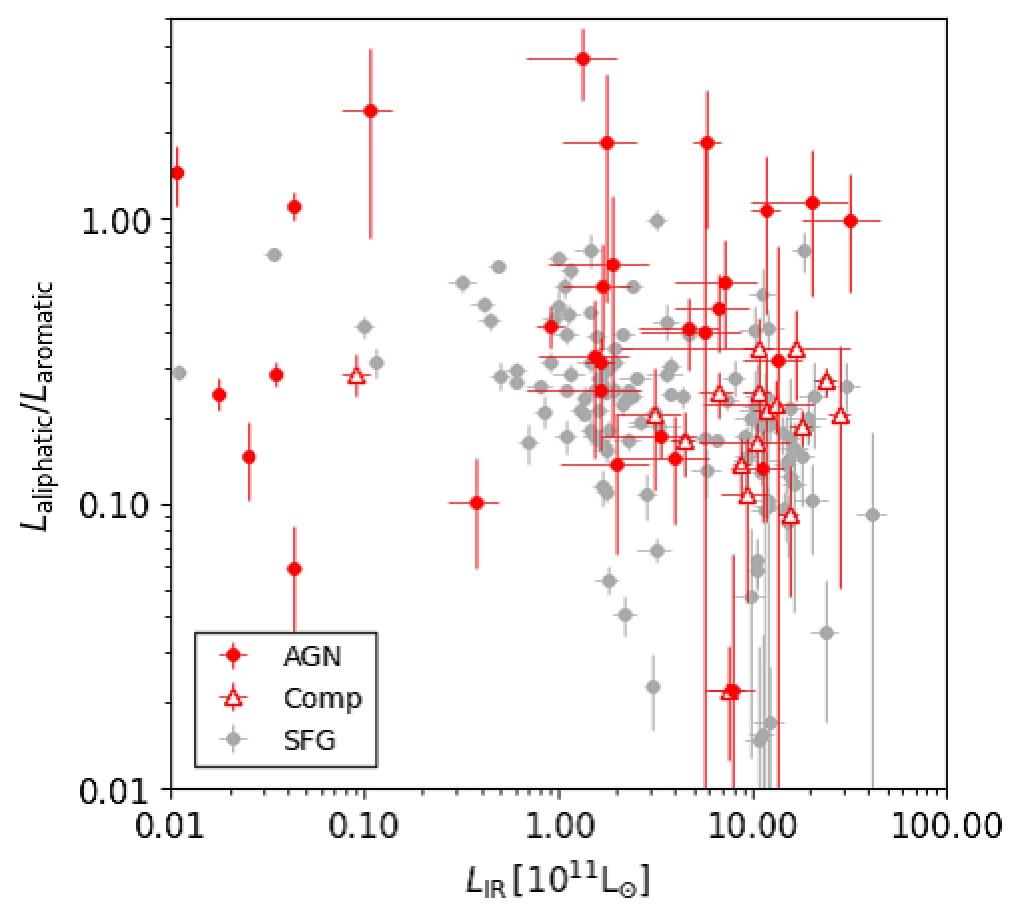} 
 \caption{Scatter plot of $L_{\rm{aliphatic}}/L_{\rm{aromatic}}$ as a function of $L_{\rm{IR}}$  
          for the AGN-dominated (red circles) and AGN-SF-composite galaxies (red triangles).
          The gray circles show the relationship for the SFG sample.}
\label{fig:lir_aliaro}
\end{figure}

We show the results related to the aliphatic-to-aromatic ratios for the aliphatic emission AGN sample.
Figure \ref{fig:lir_aliaro} displays $L_{\rm{aliphatic}}/L_{\rm{aromatic}}$ plotted as a function of $L_{\rm{IR}}$.
As can be seen in the figure, the SFG sample shows that $L_{\rm{aliphatic}}/L_{\rm{aromatic}}$ decreases with $L_{\rm{IR}}$ toward the LIRG and ULIRG classes,
 as already pointed out by \cite{Kondo2024}.
On the other hand, for the AGN sample, $L_{\rm{aliphatic}}/L_{\rm{aromatic}}$ shows no apparent dependence on $L_{\rm{IR}}$.
As a whole, we find the $L_{\rm{aliphatic}}/L_{\rm{aromatic}}$ values of the AGN sample to be systematically higher than those of the SFG sample.
The KS test indeed shows that their distributions are significantly different, with a significance level of 5$\%$.

\begin{figure}[h]
  \centering
    \includegraphics[width=7cm]{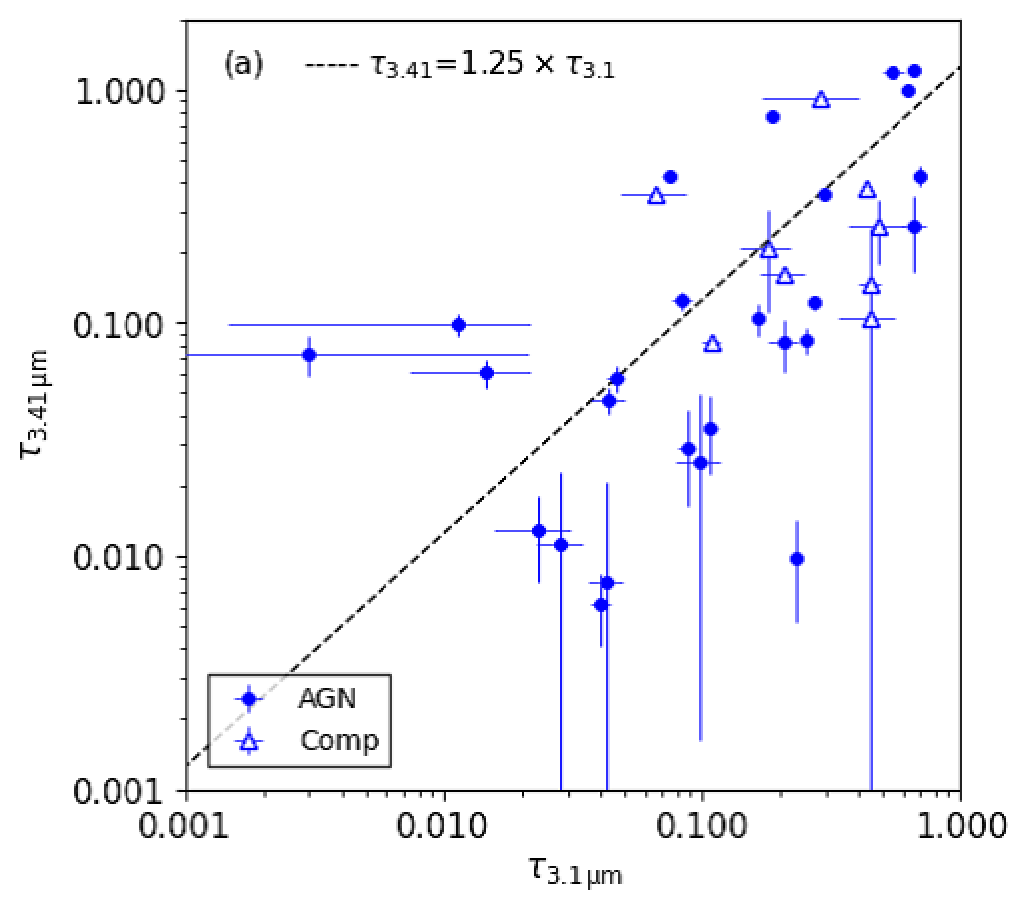} 
    \includegraphics[width=7cm]{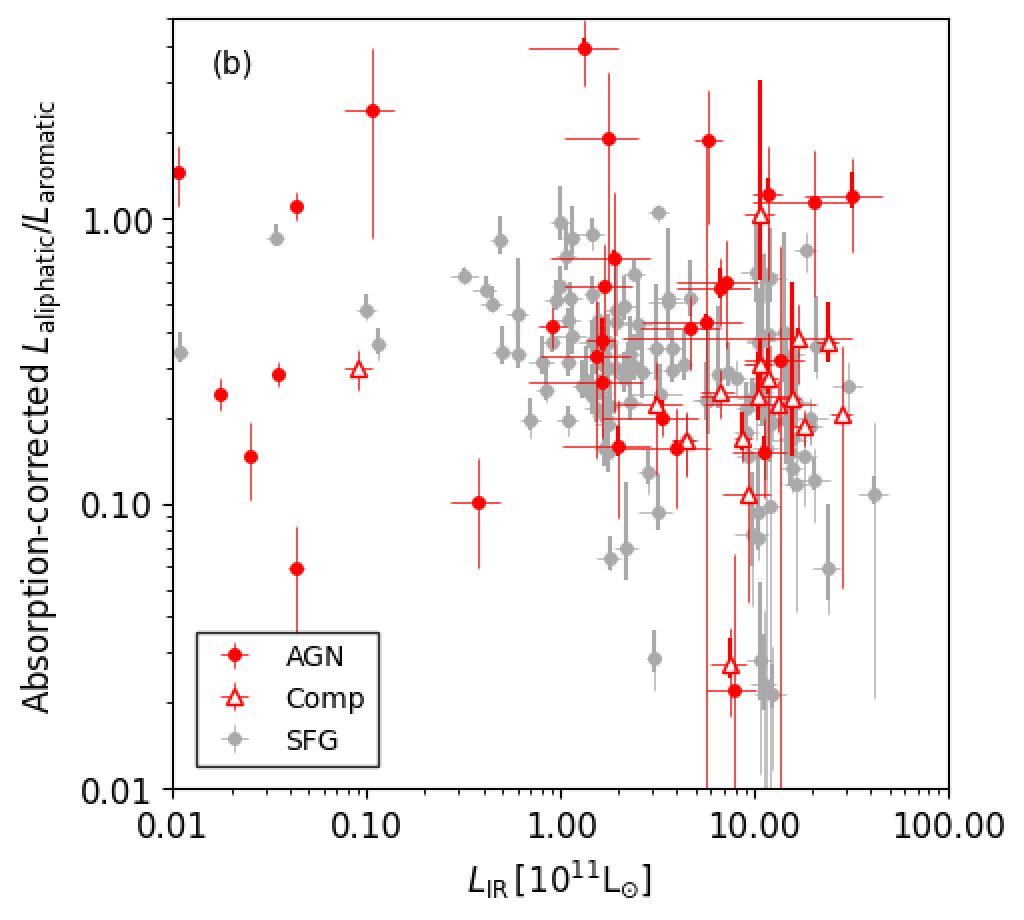} 
\caption{(a) Relationship between $\tau_{\rm{3.41}}$ and $\tau_{\rm{3.1}}$ ($\rm{H_{2}O}$ ice) 
          for the AGN-dominated (circles) and the AGN-SF-composite galaxies (triangles), both with the a-C:H absorption features.
         (b) Same as Fig. \ref{fig:lir_aliaro} but corrected for the a-C:H absorption
          that is estimated from the $\tau_{\rm{3.41}}-\tau_{\rm{3.1}}$ relation in panel (a).
          The thick error bars correspond to the systematic errors (see text for details).}
          \label{fig:lir_aliaro_corr}
\end{figure}

Since the aliphatic feature is detected in both emission and absorption, as mentioned above,
 $L_{\rm{aliphatic}}/L_{\rm{aromatic}}$ in Fig. \ref{fig:lir_aliaro} might be potentially affected by the a-C:H absorption.
Since our fitting model for the aliphatic emission sample does not consider any a-C:H absorption,
 we evaluated such an effect based on $\tau_{3.1}$ ($\rm{H_2O}$ ice), as performed in \cite{Kondo2024}.
In Fig. \ref{fig:lir_aliaro_corr}a, we plot the relationship between $\tau_{3.41}$ and $\tau_{3.1}$,
 which shows a clear correlation between them.
We corrected the $L_{\rm{aliphatic}}$ value using $\tau_{3.41}$, which is estimated 
 from its correlation with $\tau_{3.1}$ in Fig. \ref{fig:lir_aliaro_corr}a,
 where we derived the relation of $\tau_{3.41}/\tau_{3.1} = 1.25$ for the a-C:H absorption AGN sample.
Figure \ref{fig:lir_aliaro_corr}b shows $L_{\rm{aliphatic}}/L_{\rm{aromatic}}$ thus corrected for the 
a-C:H absorption plotted as a function of $L_{\rm{IR}}$,
  where the systematic errors associated with galaxy-to-galaxy variations of the relative abundance of $\rm{H_{2}O}$ ice
  are considered by assuming that the absorption correction factor can vary by a factor of two.
We confirm that the absorption effect, if any, is negligibly small.
Thus we use the uncorrected $L_{\rm{aliphatic}}$ value in the subsequent result and discussion.
\begin{figure}[h]
  \centering
    \includegraphics[width=6.4cm]{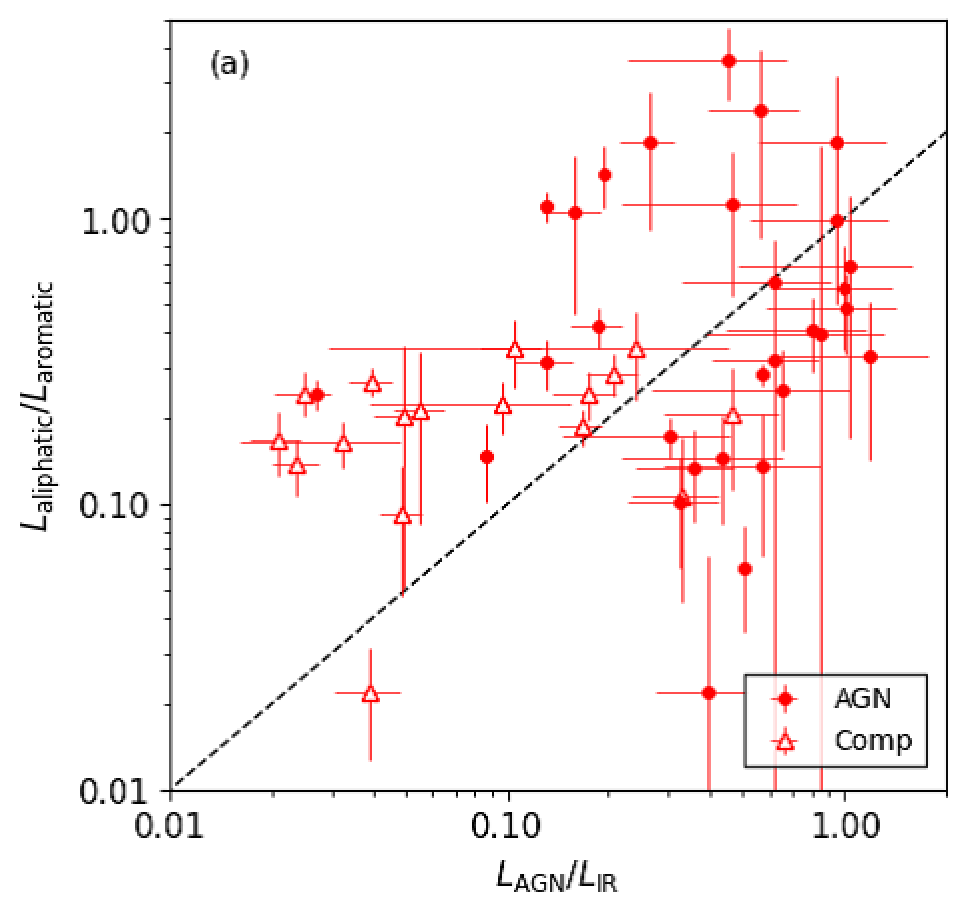} 
    \includegraphics[width=7cm]{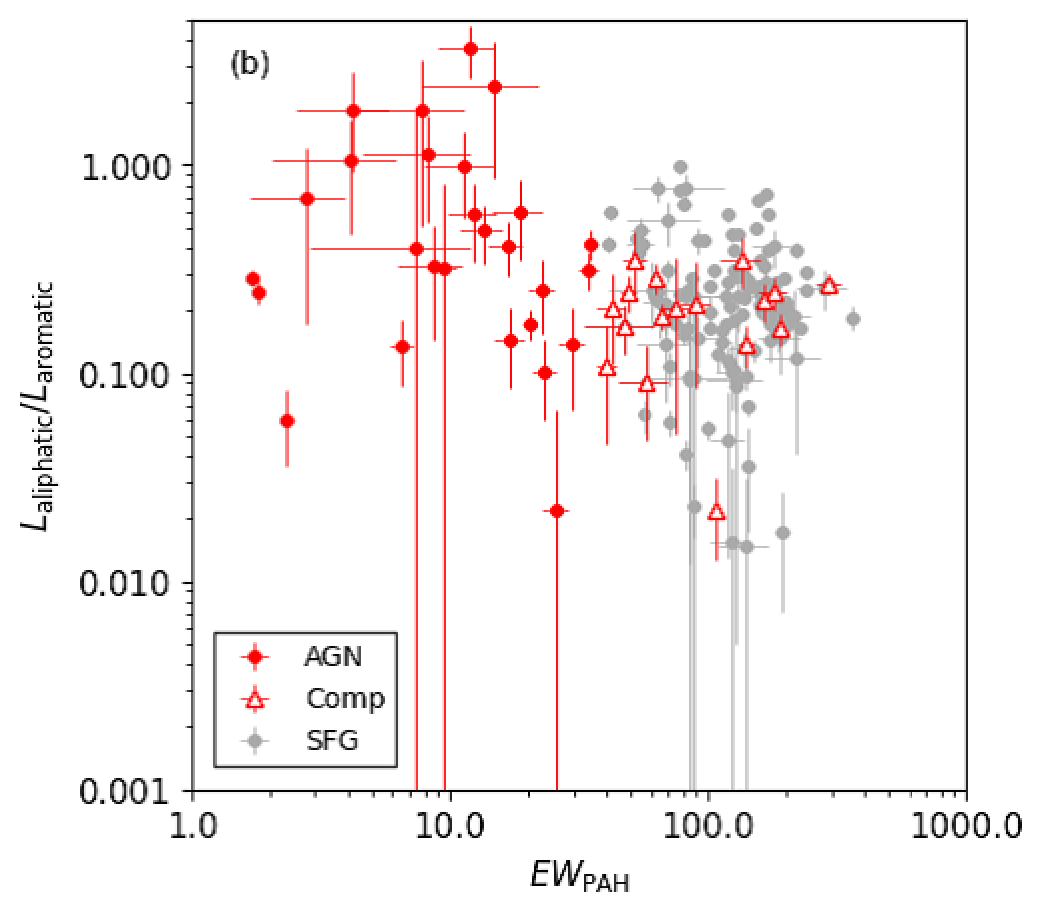} 
\caption{Scatter plots of $L_{\rm{aromatic}}/L_{\rm{aliphatic}}$ as functions of (a) $L_{\rm{AGN}}/L_{\rm{IR}}$ and (b) $EW_{\rm{PAH}}$
          for the AGN-dominated (red circles) and the AGN-SF-composite galaxies (red triangles).
          The gray circles in panel (b) show the relationship for the SFG sample.}
  \label{fig:aliaro_lhlir}
\end{figure}

\begin{figure}[h]
  \centering
    \includegraphics[width=7cm]{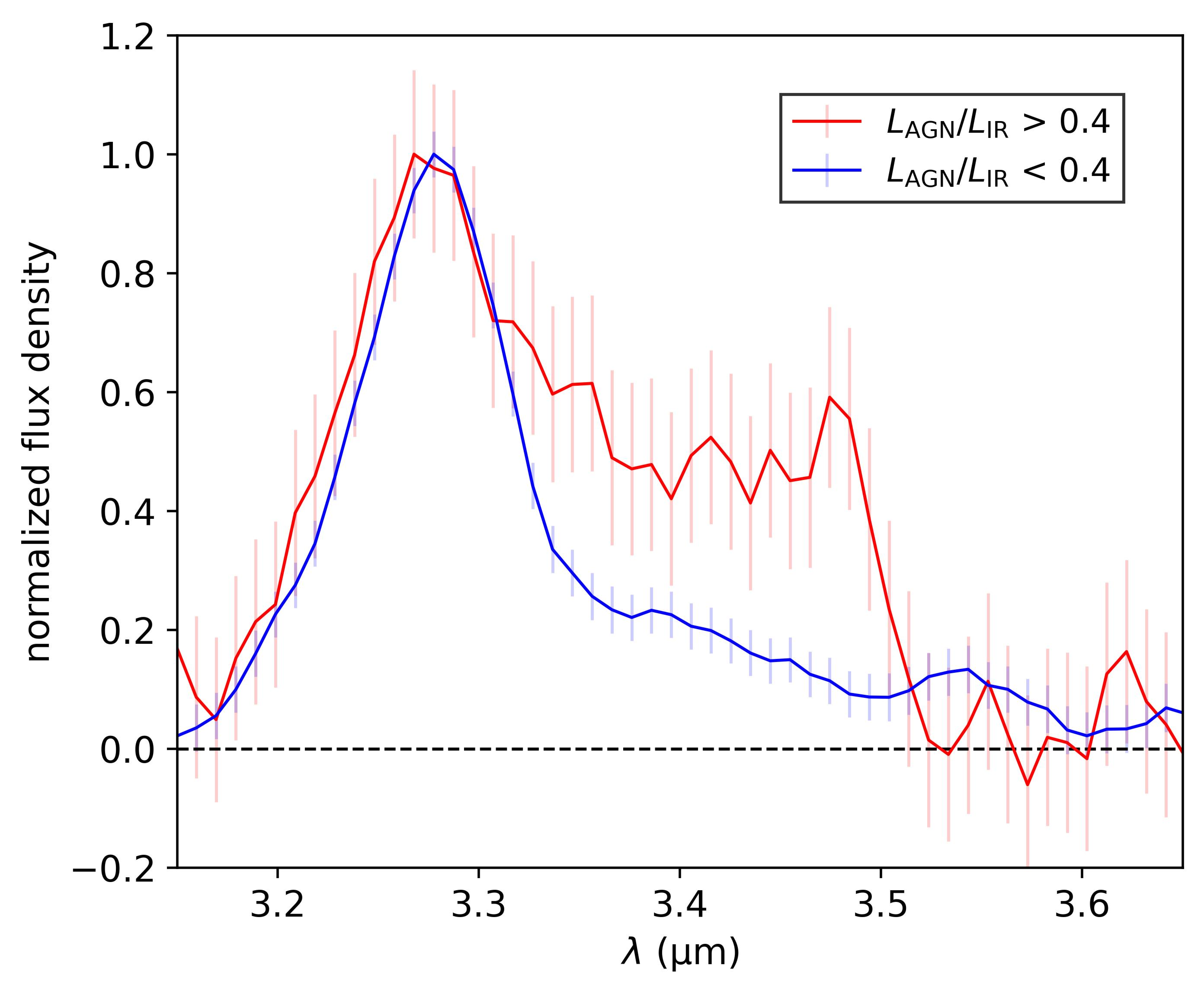} 
\caption{AKARI/IRC 3.15--3.65 $\mu$m stacked spectra of the AGN samples 
          with $L_{\rm{AGN}}/L_{\rm{IR}} > 0.4$ (red) and $L_{\rm{AGN}}/L_{\rm{IR}} < 0.4$ (blue).}
  \label{fig:stack_spec_agn}
\end{figure}

We investigated the relationship between $L_{\rm{aliphatic}}/L_{\rm{aromatic}}$ and indicators of the AGN activity.
In Fig. \ref{fig:aliaro_lhlir}a, we plot $L_{\rm{aliphatic}}/L_{\rm{aromatic}}$ as a function of $L_{\rm{AGN}}/L_{\rm{IR}}$,
from which we found that $L_{\rm{aliphatic}}/L_{\rm{aromatic}}$ significantly increases with $L_{\rm{hot}}/L_{\rm{IR}}$.
In Fig. \ref{fig:aliaro_lhlir}b, we plot $L_{\rm{aliphatic}}/L_{\rm{aromatic}}$ 
 as a function of the equivalent width of the aromatic emission feature, $EW_{\rm{PAH}}$,
 where $EW_{\rm{PAH}}$ generally takes a smaller value when the AGN activity is stronger relative to the SF activity (e.g., \citealt{Imanishi2000}).
As seen in Fig. \ref{fig:aliaro_lhlir}b, $L_{\rm{aliphatic}}/L_{\rm{aromatic}}$ decreases with $EW_{\rm{PAH}}$,
 which is consistent with the increase in $L_{\rm{aliphatic}}/L_{\rm{aromatic}}$ and thus with the AGN activity (Fig. \ref{fig:aliaro_lhlir}a).
To further visualize those results, we stacked the 3.15--3.65 $\mu$m spectra for the aliphatic emission AGN samples with $L_{\rm{AGN}}/L_{\rm{IR}}>0.4$ (red) 
and $L_{\rm{AGN}}/L_{\rm{IR}}<0.4$ (blue) separately, as shown in Fig. \ref{fig:stack_spec_agn}.
Hence the relative abundance of the aliphatic to aromatic hydrocarbons tends to increase with the AGN activity,
 which is likely to be caused by the production of aliphatic hydrocarbons 
 through some processes related to the AGN activity, as suggested in Section \ref{sect:ali_em}.

\section{Discussion}
\subsection{Properties of aliphatic hydrocarbons in AGNs\label{sect_discuss_properties}}

\begin{figure}[h]
  \centering
   \includegraphics[width=7cm]{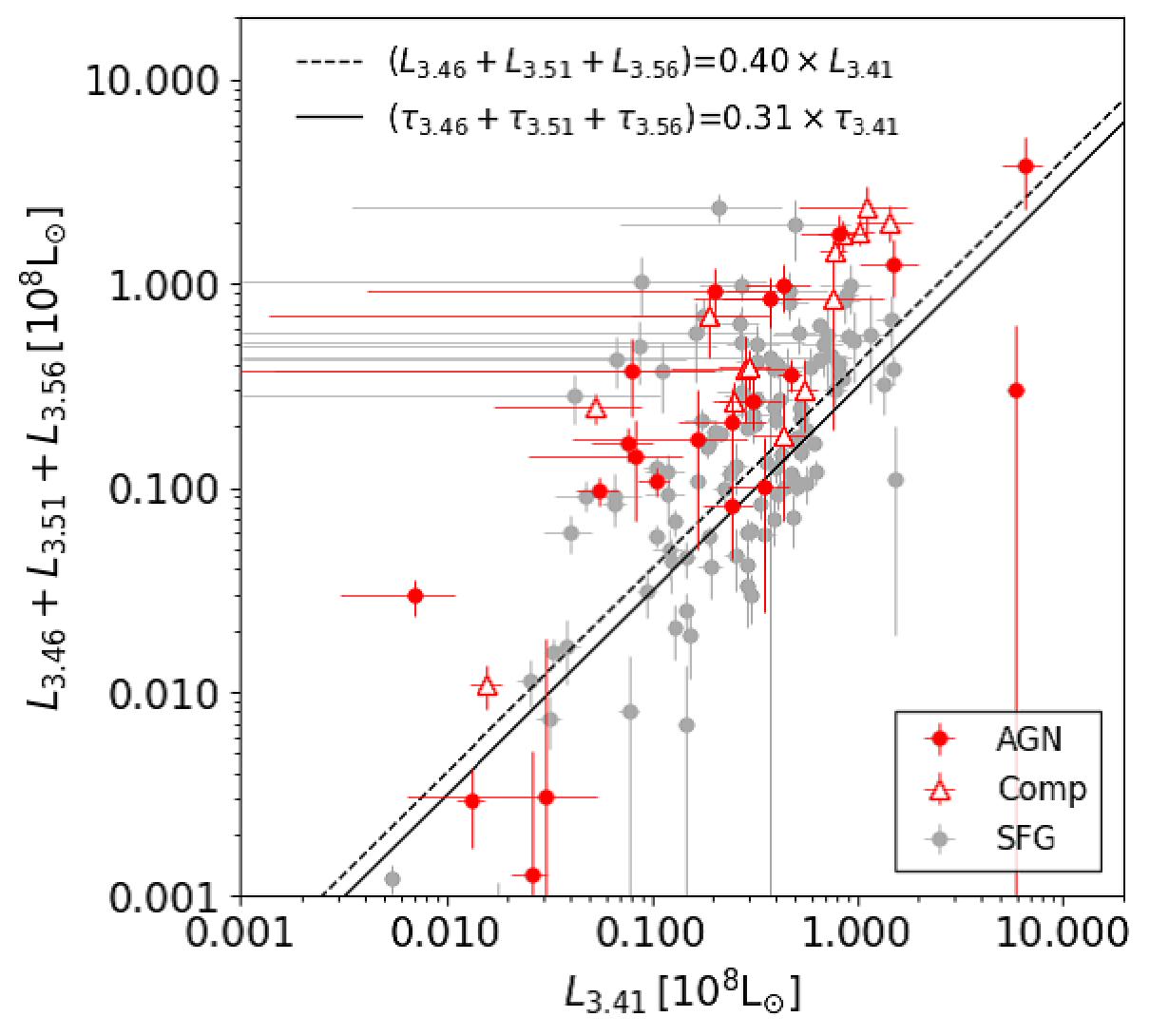} 
 \caption{Relationship between the sum of the luminosities of the three aliphatic sub-components $(L_{\rm{3.46}}+L_{\rm{3.51}}+L_{\rm{3.56}})$
           and the luminosity of the aliphatic 3.41 $\mu$m sub-component ($L_{\rm{3.41}}$) 
          for the AGN-dominated (red circles) and AGN-SF-composite galaxies (red triangles).
          The gray circles show the relationship for the SFG sample, 
           where the dashed line is the best fit to the data points.
          The solid line is the relationship between $\tau_{\rm{3.46}}+\tau_{\rm{3.51}}+\tau_{\rm{3.56}}$ and $\tau_{\rm{3.41}}$
           obtained for the stacked spectrum of the a-C:H absorption AGN sample shown in Fig. \ref{fig:stack_spec}b.
           }
 \label{fig:laliratio}
\end{figure}

\begin{figure*}[h]
  \centering
    \includegraphics[width=4.5cm,angle=90]{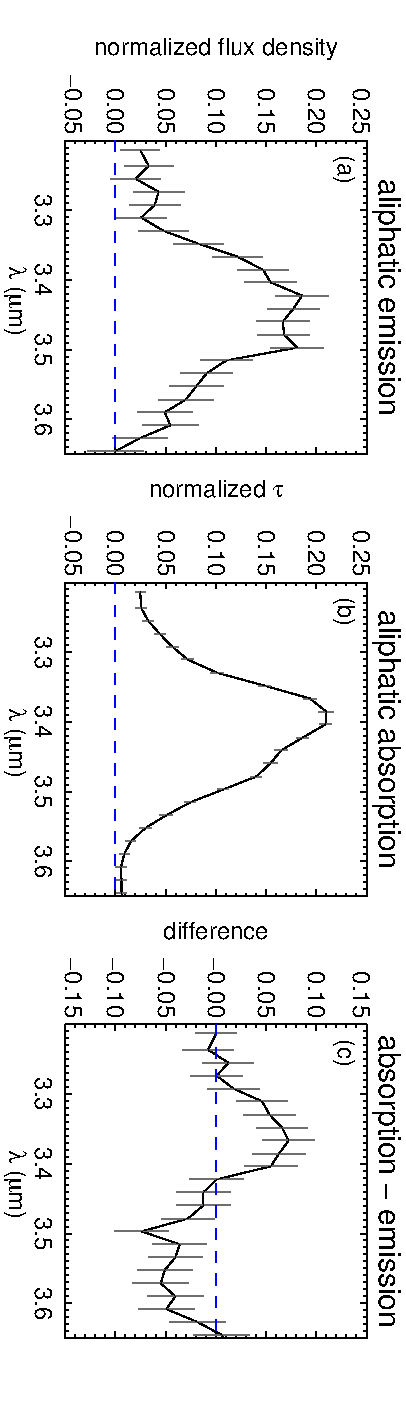} 
  \caption{AKARI/IRC 3.2--3.65 $\mu$m stacked spectra of the AGN sample. 
          (a) Aliphatic emission and (b) a-C:H absorption features, 
          where the continuum emission, the 3.3 $\mu$m aromatic hydrocarbons emission feature,
          and the $\rm{H_2O}$ ice absorption feature are removed,
          i.e., $F_{\rm{aliphatic}}$ and $\tau_{\rm{aliphatic}}$ in equations (\ref{eq:Fv_em}) and (\ref{eq:Fv_ab}).
          (c) Difference between their profiles.
          }
  \label{fig:stack_spec}
\end{figure*}

To study the properties of the aliphatic hydrocarbons in the AGN sample,
 we investigated the relationship between the relative strengths of the four spectral sub-components characterizing the aliphatic feature.
 Considering a balance of the statistical significance,
 we added the fluxes of the three longer wavelength sub-components (3.46, 3.51, and 3.56 $\mu$m) 
 and obtained their relationship with the 3.41 $\mu$m sub-component.
In Fig. \ref{fig:laliratio}, we plot the relationship between the sum of the luminosities of the three aliphatic sub-components $(L_{\rm{3.46}}+L_{\rm{3.51}}+L_{\rm{3.56}})$
 and the luminosity of the aliphatic 3.41 $\mu$m sub-component ($L_{\rm{3.41}}$).
To estimate the flux errors of $L_{\rm{3.46}}+L_{\rm{3.51}}+L_{\rm{3.56}}$ and $L_{\rm{3.41}}$,
 we fixed the relative strengths of the three longer wavelength sub-components at the best-fit values and refit the spectra.
Figure \ref{fig:laliratio} shows the results of both AGN and SFG samples,
 where the best-fit relation for the SFG sample is indicated by the dashed line for comparative purposes.
The figure suggests that the profiles of the aliphatic emission features of the AGN sample 
 may be different from those of the SFG sample in that the feature intensities of the AGN sample are systematically stronger at longer wavelengths.

To visualize the shape of the aliphatic feature more robustly, we stacked the 3.2--3.65 $\mu$m spectra
 for the aliphatic emission AGN sample and the a-C:H absorption AGN sample separately.
For the purpose of comparison, as shown in Fig. \ref{fig:stack_spec}, the flux density and the optical depth,
 which are given in arbitrary units, are scaled so that the integral of each of the emission and the absorption feature
 with respect to the frequency has the same area with each other.
In the figure, we recognize a clear difference in the profile of the aliphatic feature between the emission and the absorption:
 The 3.41 $\mu$m sub-component is weaker relative to the longer wavelength sub-components in the emission feature than in the absorption feature.
Figure \ref{fig:stack_spec}c more clearly demonstrates such a significant difference in the properties of the hydrocarbons
 between the aliphatic emission and the a-C:H absorption feature.
We characterized the stacked absorption feature in Fig. \ref{fig:stack_spec}b with the absorption model described in Section \ref{sect_spec_fit}
 to obtain the relation of $\tau_{\rm{3.46}}+\tau_{\rm{3.51}}+\tau_{\rm{3.56}}=0.31\times\tau_{\rm{3.41}}$,
 which is much closer to the relationship between $L_{\rm{3.46}}+L_{\rm{3.51}}+L_{\rm{3.56}}$ and $L_{\rm{3.41}}$ for the SFG sample than for the AGN sample,
 as indicated by the solid line in Fig. \ref{fig:laliratio}.
Hence, even within the AGN sample, the properties of aliphatic hydrocarbons are different between those emitting and absorbing,
 the latter of which are thought to exist in the interstellar space of host galaxies (Sections \ref{sect_result_aro} and \ref{sect_origin_aliaro}),
 and thus it is reasonable that their properties are similar to those of the SFG sample.

The relative strengths of the sub-components characterizing the aliphatic feature are expected to change,
 depending on differences in the stretching mode and the chemical composition of the hydrocarbons.
The aliphatic feature at 3.4--3.6 $\mu$m consists of both symmetric and asymmetric stretching modes,
 where the asymmetric modes exhibit the peak at shorter wavelengths than the symmetric modes \citep{Kwok2012}. 
Regarding the chemical composition relevant to the C$-$H vibration,
 the spectral feature of methyl ($\rm{-CH_3}$) groups have peaks at shorter wavelengths,
 while those of methylene ($\rm{-CH_2}$) groups and C$-$H groups have peaks at longer wavelengths \citep{Kwok2012}.
Therefore, the difference in the profile of the aliphatic emission feature relative to the a-C:H absorption feature 
 favors the conditions that the emitting hydrocarbons are less excited in the asymmetric stretching modes 
 and more C-rich in chemical composition.
The former suggests that a typical temperature of the hydrocarbon dust may be lower, and thus its typical size may be larger,
 while the latter implies that hydrogenated amorphous carbons, for instance, are more likely than chain-like hydrocarbons to be representatives of the emitting aliphatic feature carriers.

\subsection{Possible origins of aromatic and aliphatic hydrocarbons in AGNs\label{sect_origin_aliaro}}

\begin{figure}[h]
  \centering
   \includegraphics[width=7cm,clip]{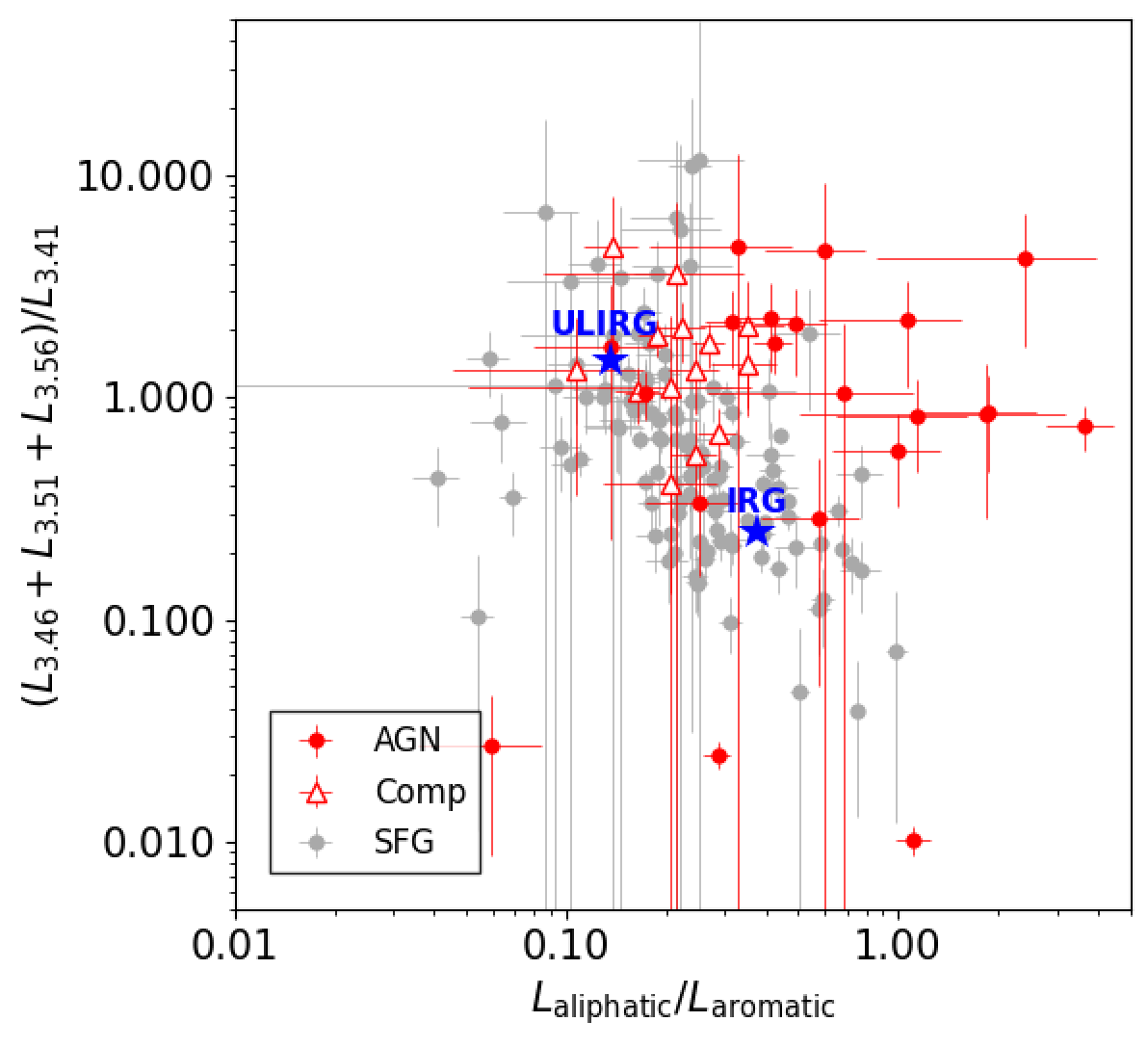} 
   \caption{Scatter plot of the relative strength of the aliphatic sub-components $(L_{\rm{3.46}}+L_{\rm{3.51}}+L_{\rm{3.56}})/L_{\rm{3.41}}$
             as a function of $L_{\mathrm{aliphatic}}/L_{\mathrm{aromatic}}$
            for the AGN-dominated (red circles) and AGN-SF-composite galaxies (red triangles).
            The gray circles show the relationship for the SFG sample.
            The blue star marks show the mean values of the IRGs and the ULIRGs for the SFG sample.}
   \label{fig_aliaro_aliratio}
  \end{figure}

As described in Section \ref{sect_result_aro}, $L_{\rm{aromatic}}/L_{\rm{IR}}$ decreases with $\tau_{3.41}$,
 indicating that the aromatic emission is affected by the a-C:H absorption and thus the interstellar dust extinction.
To confirm the validity of this picture, we corrected $L_{\rm{aromatic}}/L_{\rm{IR}}$ using the standard extinction curve.
From the near-IR continuum extinction ($A_{\rm{L}}/A_{\rm{V}} = 0.056$; \citealt{Cardelli1989}) and 
 the 3.4 $\mu$m optical depth relative to the visual extinction ($A_{\rm{V}}/\tau_{3.4} \approx 300$ in the local diffuse ISM; \citealt{Gao2010}),
 we estimated the optical depth of the near-IR continuum around the aromatic feature, $\tau_{\rm{cont}}$, to be $\tau_{\rm{cont}} \approx 20\times \tau_{3.4}$.
Assuming that an optical depth typical of ULIRGs is $\tau_{3.41}\approx0.2$ (Fig. \ref{fig:lir_aro_ab}b),
 the extinction-corrected $L_{\rm{aromatic}}/L_{\rm{IR}}$ is estimated to be 30 times larger 
 and thus becomes comparable to the $L_{\rm{aromatic}}/L_{\rm{IR}}$ of the aliphatic emission AGN sample.
Therefore, in line with the results of previous studies (e.g., \citealt{Alonso-Herrero2014}; \citealt{Jensen2017}; \citealt{Garcia2022b}), 
 the aromatic hydrocarbons likely exist in the circumnuclear region and are
 shielded from the hard radiation field by the dense gas surrounding the AGNs, 
 and their emission is attenuated significantly by the interstellar dust extinction in the host galaxy.
Moreover, as seen in Fig. \ref{fig:lir_aro_ab}b, galaxies classified as AGN-SF-composite have a $\tau_{3.41}$ significantly larger than
AGN-dominated galaxies at high luminosities.
Given that many ULIRGs are known to host CONs (e.g., \citealt{Falstad2021}),
 most of them may belong to AGNs with the CONs and not be significantly affected by the AGN activity.
In this context, after revisiting Fig. \ref{fig:lir_aro_ab}a, \ref{fig:lir_ali}a, and \ref{fig:lir_aliaro},
 we find that the AGN-SF-composite ULIRGs show the properties of the aromatic and aliphatic emission features
 similar to those of the SFG sample, thus making it reasonable to interpret that the PAHs are shielded by dense gas surrounding CONs.

In the following, we discuss the origin of the aliphatic hydrocarbons in AGNs.
As described in Section \ref{sect:ali_em}, the aliphatic emission intensities have an increasing trend with the AGN activity,
 suggesting that the aliphatic hydrocarbons may exist within the reach of the AGN activity and thus in the circumnuclear region, similar to the aromatic hydrocarbons.
If the emission features of the aliphatic and aromatic hydrocarbons were of the same carrier origins, 
 such as mixed aromatic-aliphatic organic nanoparticles (\citealt{Kwok2011,KwokZhang2013}),
 the aliphatic hydrocarbons would likely be removed more easily since the aliphatic bonds are chemically weaker than the aromatic ones.
Indeed, the JWST observation of the Seyfert galaxy NGC 7469 shows that the aliphatic-to-aromatic ratio becomes lower near the nucleus \citep{Lai2023},
and some of our AGNs also show a relatively low $L_{\rm{aliphatic}}/L_{\rm{aromatic}}$, especially at low luminosities.
On the contrary, for a significant fraction of the other AGNs,
$L_{\rm{aliphatic}}/L_{\rm{aromatic}}$ increases with the AGN activity as shown in Fig. \ref{fig:aliaro_lhlir},
 and thus the aliphatic hydrocarbons present in the circumnuclear region are likely to be of different carrier origins from the aromatic hydrocarbons.
Those aliphatic hydrocarbons may come from a new population created 
 through processes such as shattering of aliphatic-rich amorphous carbon dust (e.g., \citealt{Jones2013}) by AGN outflows.
For example, \cite{Yamagishi2012} suggested the production of small carbonaceous dust 
 through shattering of larger carbonaceous grains in the harsh galactic halo of M82.
However, \cite{Jones2013} suggested rather aromatic-rich carbonaceous dust surfaces processed by the interstellar UV field,
as in the present case where the hard radiation field can be suppressed by the dense gas surrounding the AGNs,
the carbonaceous dust may mostly remain aliphatic rich.

Finally, we discuss how the likely new dust population may be related to 
 the unusual properties of the aliphatic emission feature in the AGN sample as described in Section \ref{sect_discuss_properties}.
Figure \ref{fig_aliaro_aliratio} shows the scatter plot of the relative strengths of 
 the aliphatic sub-components $(L_{\rm{3.46}}+L_{\rm{3.51}}+L_{\rm{3.56}})/L_{\rm{3.41}}$ as a function of $L_{\mathrm{aliphatic}}/L_{\mathrm{aromatic}}$,
 from which we find that the distribution of the data points of the AGN sample is well separated from that of the SFG sample.
Regarding the SFG sample, our new finding, which was not discussed in \cite{Kondo2024},
 is that the properties of the aliphatic emission features, $(L_{\rm{3.46}}+L_{\rm{3.51}}+L_{\rm{3.56}})/L_{\rm{3.41}}$,
 clearly change with the decreasing $L_{\mathrm{aliphatic}}/L_{\mathrm{aromatic}}$ from IRG to ULIRGs, as seen in Fig. \ref{fig_aliaro_aliratio}. 
This may suggest that the small H-rich aliphatic hydrocarbons are more easily photo-dissociated under the strong radiation fields associated with active SF,
 resulting in the dominance of relatively large C-rich aliphatic hydrocarbons in such harsh interstellar conditions.
On the other hand, for the AGN sample, $(L_{\rm{3.46}}+L_{\rm{3.51}}+L_{\rm{3.56}})/L_{\rm{3.41}}$ stays at a level as high as the ULIRGs of the SFG sample,
 while $L_{\mathrm{aliphatic}}/L_{\mathrm{aromatic}}$ extends to ratios higher than those of the IRGs,
 suggesting that the new population of relatively large-sized aliphatic hydrocarbon dust created in the circumnuclear region increases $L_{\mathrm{aliphatic}}/L_{\mathrm{aromatic}}$, 
 which may possess C-rich chemical compositions intrinsically, as discussed in Section \ref{sect_discuss_properties}.

\section{Conclusions}
We have conducted spectral fitting and SED fitting for our sample of 102 AGNs 
 in order to estimate the properties of the aromatic and aliphatic hydrocarbon dust under the influence of AGN activity.
For the aromatic hydrocarbons, 65 AGNs are detected at above 5$\sigma$. 
The $L_{\mathrm{aromatic}}/L_{\mathrm{IR}}$ values of the AGN sample are systematically lower than those of the SFG sample
 and much lower for the a-C:H absorption AGN sample,
 suggesting that their emission is significantly suppressed due to the interstellar dust extinction in the outer regions of the host galaxy.
On the other hand, the $L_{\mathrm{aliphatic}}/L_{\mathrm{aromatic}}$ values of the AGN sample are systematically higher than those of the SFG sample.
Regarding the relationship with the AGN activity, the
 $L_{\mathrm{aliphatic}}/L_{\mathrm{aromatic}}$ of the AGN sample shows an increasing trend 
 with the fractional luminosity of the AGN component, $L_{\rm{AGN}}/L_{\rm{IR}}$.
The overall results indicate that both aromatic and aliphatic hydrocarbon dust may be of circumnuclear origins;
 in particular, a significant fraction of the observed aliphatic hydrocarbon dust is likely to be newly produced 
 due to interaction through processes such as the shattering of large carbonaceous grains by AGN outflow, 
 which can explain the high $L_{\mathrm{aliphatic}}/L_{\mathrm{aromatic}}$ of the AGN sample.

Furthermore, by comparing the profiles of the aliphatic features of the aliphatic emission and the absorption AGN samples,
 we found a clear difference in the shapes of their profiles.
 The longer-wavelength sub-components are relatively strong in the emission feature,
 while the 3.41 $\mu$m sub-component is relatively strong in the absorption feature, which is similar to the profile
 usually seen in SFGs of $L_{\rm{IR}}$ lower than $10^{12}\,\rm{L_{\odot}}$. The newly produced aliphatic hydrocarbons in AGNs may be attributed to the unusual profiles of the aliphatic emission feature.

 \begin{acknowledgements}
  We thank the referee for giving us useful comments.
  This research is based on observations with AKARI, a JAXA project with the participation of ESA. 
  This publication makes use of data products from the Wide-field Infrared Survey Explorer, 
  which is a joint project of the University of California, Los Angeles, 
  and the Jet Propulsion Laboratory/California Institute of Technology, funded by the National Aeronautics
  and Space Administration, and data products from the Infrared Astronomical Satellite (IRAS), 
  which is a joint project of the US, UK and the Netherlands. 
  Spitzer Space Telescope is operated by the California Institute of Technology for NASA under NASA contract 1407. 
 \end{acknowledgements}
 
\bibliography{ref_paper}
\bibliographystyle{aasjournal}
\end{document}